\documentclass[twocolumn]{aastex62}

\graphicspath{{./}{figures/}}

\received{}
\revised{}
\accepted{}
\submitjournal{ApJ}

\shorttitle{SF quenching since z\,$\sim$\,1}
\shortauthors{Pintos-Castro et al.}

\usepackage{amsmath,amstext}
\usepackage{multirow}

\begin{document}

\title{THE EVOLUTION OF THE QUENCHING OF STAR FORMATION IN CLUSTER GALAXIES SINCE Z\,$\sim$\,1}

\email{pintos@astro.utoronto.ca}

\author[0000-0002-9133-4457]{I. Pintos-Castro}
\affiliation{Department of Astronomy \& Astrophysics,
University of Toronto, 50 St. George Street, 
Toronto, ON M5S 3H4, Canada}
\author{H. K. C. Yee}
\affiliation{Department of Astronomy \& Astrophysics,
University of Toronto, 50 St. George Street,
Toronto, ON M5S 3H4, Canada}
\author{A. Muzzin}
\affiliation{Department of Physics and Astronomy, York University, 4700 Keele Street, Toronto, ON M3J 1P3, Canada}
\author{L. Old}
\affiliation{Department of Astronomy \& Astrophysics,
University of Toronto, 50 St. George Street,
Toronto, ON M5S 3H4, Canada}
\affiliation{European Space Astronomy Centre (ESAC), Villanueva
de la Ca\~nada, E-28692 Madrid, Spain}
\author[0000-0002-6572-7089]{G. Wilson}
\affiliation{Department of Physics and Astronomy, University of California, Riverside, CA 92521, USA}

\begin{abstract}

We study the star-forming (SF) population of galaxies within a sample of 209 IR-selected galaxy clusters at 0.3$\,\leq\,z\,\leq\,$1.1 in the ELAIS-N1 and XMM-LSS fields, exploiting the first HSC-SSP data release. The large area and depth of these data allows us to analyze the dependence of the SF fraction, $f_{SF}$, on stellar mass and environment separately. Using $R/R_{200}$ to trace environment, we observe a decrease in $f_{SF}$ from the field towards the cluster core, which strongly depends on stellar mass and redshift. The data show an accelerated growth of the quiescent population within the cluster environment: the $f_{SF}$ vs. stellar mass relation of the cluster core ($R/R_{200}\,\leq\,$0.4) is always below  that of the field (4$\,\leq\,R/R_{200}\,<\,$6). 
Finally, we find that environmental and mass quenching efficiencies depend on galaxy stellar mass and distance to the center of the cluster, demonstrating that the two effects are not separable in the cluster environment. We suggest that the increase of the mass quenching efficiency in the cluster core may emerge from an initial population of galaxies formed ``in situ.''
The dependence of the environmental quenching efficiency on stellar mass favors models in which galaxies exhaust their reservoir of gas through star formation and outflows, after new gas supply is truncated when galaxies enter the cluster.

\end{abstract}


\section{Introduction}
Observationally, galaxies can be classified into two broad types: red passive galaxies dominated by early-type morphologies and blue star-forming galaxies dominated by late-type morphologies \citep[e.g.,][]{Strateva2001,Baldry2004}. The probability of a galaxy belonging to one of these types is mainly driven by its stellar mass and its surrounding environment \citep[e.g.,][]{Dressler1980,Kauffmann2004,Balogh2004,Baldry2006}: on average, more massive galaxies are more likely to become quiescent, and galaxies in denser regions are more likely to become quenched. Another important parameter is the redshift: while local clusters are populated by red and dead massive galaxies, the fraction of blue, star-forming galaxies is higher in more distant clusters \citep[e.g.,][]{Butcher1984,Li2009,Raichoor2012,Hennig2017}, and there is even evidence for high star formation activity close to the cluster core of the highest redshift galaxy clusters \citep[e.g.,][]{Hilton2010,Santos2015,Webb2015,Wang2016}.

Several studies have focused on disentangling the separate contributions of processes driven by intrinsic galaxy properties and the environmental influence. It is the ``nature vs. nurture'' scenario of the debate \citep{Dressler1980}. Over the last decade, a number of investigations have classified mechanisms of suppression of star formation activity into two major categories known as ``mass quenching,'' related to internal processes that scale with stellar mass, and ``environmental quenching,'' related to external changes as galaxies interact with their surroundings \citep{Peng2010}. Whether these two quenching processes act independently from each other is still under discussion. While observational studies show evidence that the two effects are separable in the local Universe or for relatively high-mass galaxies at high redshifts \citep{Peng2010,Quadri2012,Muzzin2012,Kovac2014}; there is growing observational evidence that environmental quenching evolves and depends on stellar mass \citep{Balogh2016,Darvish2016,Kawinwanichakij2017,Lin2017temp}. 

Some of the apparent discrepancies when studying ``environment'' may arise from its many different definitions. One approach is to analyze galaxy properties in terms of the large scale structure in which galaxies are located (e.g., super-clusters, clusters, groups, field, voids). In particular, a commonly used proxy is the projected distance to the center of the parent halo to find at which radius galaxy properties are effectively altered \citep[e.g.,][]{Wetzel2014,Allen2016,Vulcani2016,Bufanda2017}. Another useful approach, which requires spectroscopic data to build up a picture of the phase space profiles of clusters, is to identify whether galaxies are infalling or virialized to determine how long they have been under the influence of the host halo \citep{Muzzin2014,Jaffe2015,Jaffe2018,Noble2016,Weinzirl2017}.  A common approach is to study galaxy properties in terms of the local galaxy density around each individual galaxy: in this case no previous knowledge of clusters/groups positions are needed \citep[e.g.,][]{Peng2010,Li2012,Davies2016,Kawinwanichakij2017,Darvish2018}. There is no universal method to measure the environment and its suitability depends on the physical scale being probed \citep[see e.g.,][for a detailed discussion on galaxy environment measures]{Muldrew2012}.

Physical explanations for the suppression of star formation in galaxies related to mass include feedback from an active galactic nucleus \citep[e.g.,][]{Fabian2012} or energetic feedback from supernova explosions and stellar winds \citep[e.g.,][]{Dekel1986,Veilleux2005}. Regarding the environmental influence, many physical processes have been proposed which might transform the morphology and star formation properties of galaxies in dense environments, acting at different timescales and physical distances from the center of the host halo \citep[see, e.g.,][for a review]{Boselli2006}. Such mechanisms can be broadly classified into: (1) the interaction of a galaxy with the gaseous component of the cluster \citep[e.g., ram-pressure stripping;][]{Gunn1972}; (2) interactions between the galaxy and the cluster gravitational potential \citep[e.g., tidal stripping;][]{Merritt1984}; and, (3) smaller scale galaxy-galaxy interactions \citep[e.g., harassment;][]{Moore1996}. The relative importance of quenching mechanisms also depends on the epoch. For example, at high redshift, in the absence of cosmological accretion, once a galaxy falls into a larger halo \citep{Dekel2006}, vigorous star formation may drive outflows that will exhaust the gas supply in short timescales \citep[``overconsumption'', see][]{McGee2014,Balogh2016}.

To have a clearer picture of the quenching mechanisms in play, one needs to track how galaxies change their star formation properties as they fall into the cluster, isolating the role of the environment from galaxies' internal processes. In this paper, we combine the SpARCS \citep[Spitzer Adaptation of the Red-Sequence Cluster Survey,][]{Muzzin2009,Wilson2009} sample of clusters with the first public data release from the HSC-SSP \citep[Hiper Suprime-Cam Subaru Strategic Program][]{Aihara2017} in the XMM-LSS and ELAIS-N1 fields. We focus on examining the fraction of star-forming cluster galaxies, out to as far as 10 virial radii, with a sample sufficiently large enough to perform background corrections, to control for the galaxies intrinsic properties (such as stellar mass), over a large range of redshift. This allows us to track how star formation is quenched as galaxies in-fall into clusters, providing a clearer picture of how galaxies in clusters evolve.

The paper is structured as follows. Section \S\,2 introduces the data set, along with the cluster galaxies selection. Section \S\,3 describes how galaxy properties have been determined, including possible systematics. In section \S\,4 we present our results, focusing on the dependence of the fraction of star-forming galaxies with stellar mass and distance to the cluster center at different redshift bins. We also quantify the environmental and mass quenching efficiencies and their dependences on stellar mass and cluster-centric radius. In Section \S\,5 we discuss our results and in \S\,6 a summary of this work is given. Unless otherwise stated, throughout this paper we adopt a $\Lambda$CDM cosmology with $H_{0}$\,=\,70\,km\,s$^{-1}$\,Mpc$^{-1}$, $\Omega_{m}$\,=\,0.3, and $\Omega_{\Lambda}$\,=\,0.7. All magnitudes are on the AB system.

\begin{figure}
\centering
\includegraphics[width=\hsize]{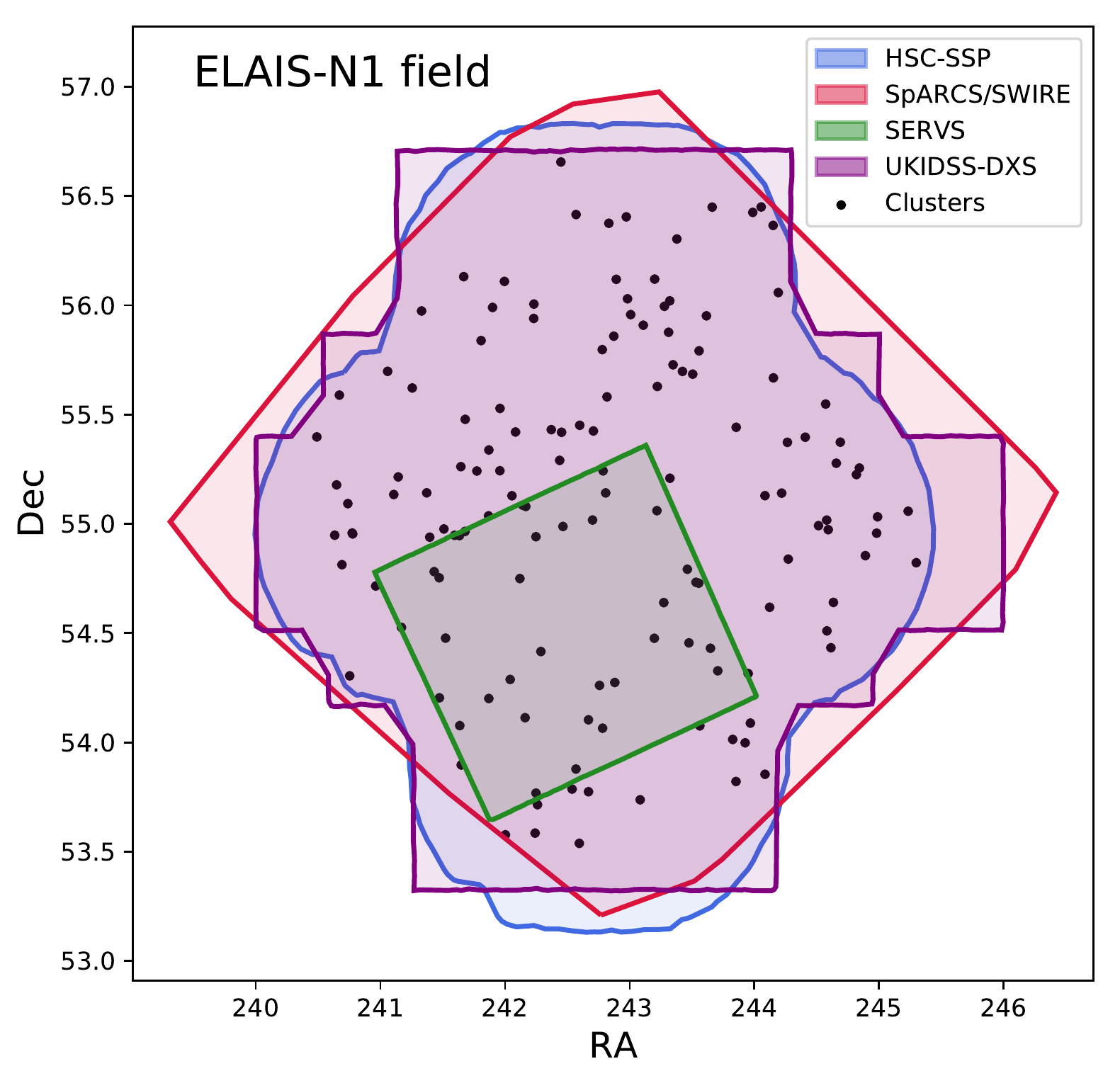}
\includegraphics[width=\hsize]{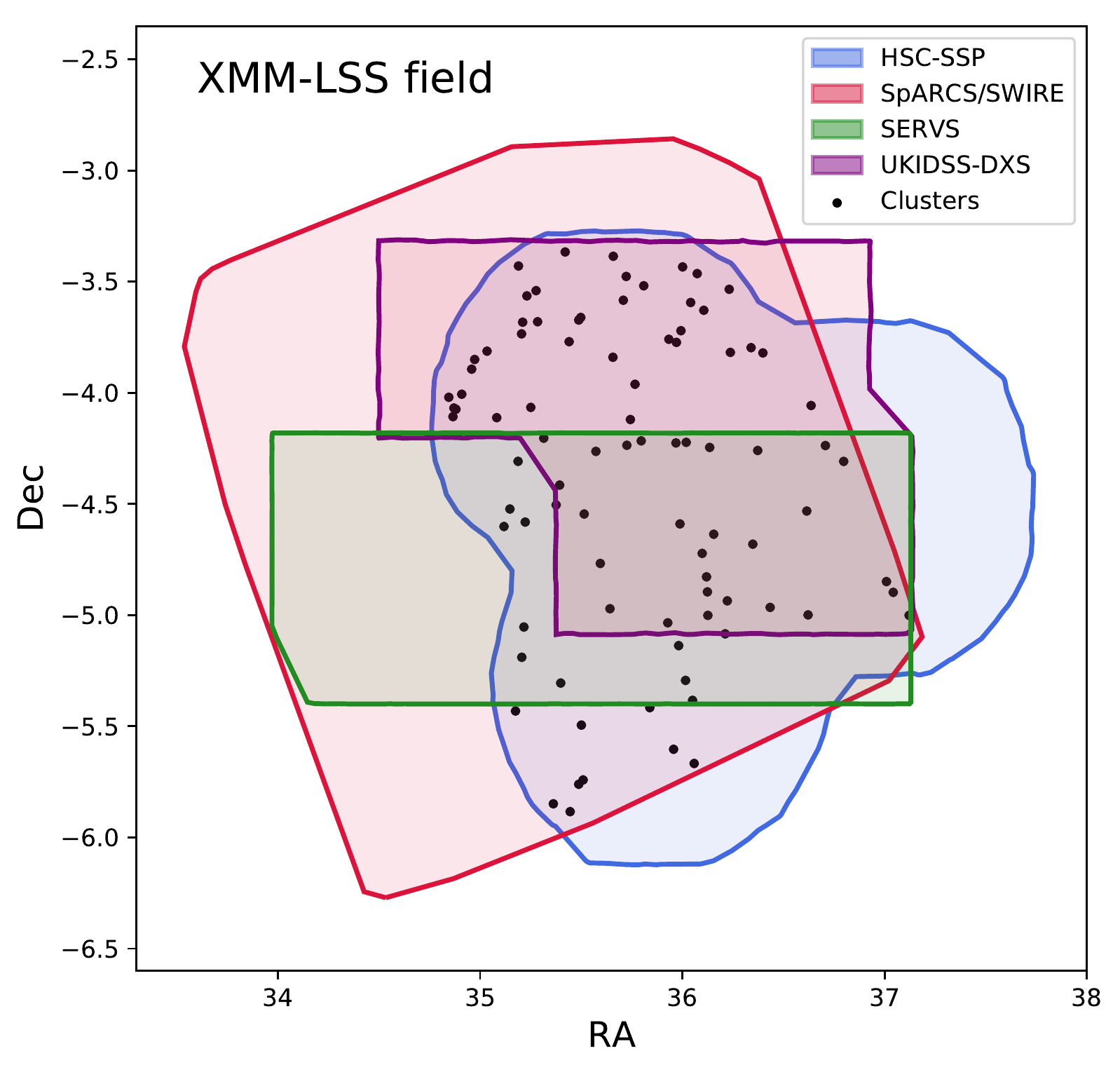}
\caption{Coverage of the available data in the ELAIS-N1 (top) and XMM-LSS (bottom) fields used in the present work. Blue regions represent the HSC-SSP deep layer from the first data release; red areas show the SpARCS/SWIRE coverage; green areas are the SERVS data; and, purple regions come from UKIDSS-DSX. The clusters analyzed in this study (black circles, see \S\,\ref{sec:cl_sample}) are in the overlapping regions of HSC and SpARCS.}
\label{fig:cat_coverage}
\end{figure}


\section{Data and sample selection}
\subsection{Optical and infrared data}

We use the first public data release from the HSC-SSP\footnote{\url{http://hsc.mtk.nao.ac.jp/ssp/}} \citep{Aihara2017} project for the optical broad-band photometry: $g$, $r$, $i$, $z^{\prime}$, and $Y$ bands. This catalog covers a large region of the XMM-LSS SpARCS field and the majority of the ELAIS-N1 SpARCS field, as shown in Figure \ref{fig:cat_coverage}. The total overlapping area is about 16.5\,deg$^{2}$. The HSC-SSP survey consists of three layers of depth (wide, deep, and ultra-deep), with our fields of interest being part of the deep layer. The 5\,$\sigma$ depths estimated by the HSC-SSP team for the deep layer are 26.8, 26.6, 26.5, 25.6, and 24.8\,mag for the $g$, $r$, $i$, $z$, and $Y$ bands, respectively.

For NIR data we benefit from the UKIDSS project \citep{Lawrence2007}. 
From the final data release of DXS (Deep Extragalactic Survey, DR10), we have J and K data covering a significant portion of both fields. 

Additional NIR photometry comes from the SWIRE Legacy survey \citep{Lonsdale2003}, which is the NIR band photometry for SpARCS. This survey has imaged nearly 50 deg$^2$, in six high-latitude fields, with \textit{Spitzer} at 3.6, 4.5, 5.8, and 8\,$\mu$m using IRAC \citep{Fazio2004} and at 24\,$\mu$m using MIPS \citep{Rieke2004}. Additional deeper observations in IRAC\,3.6 and 4.5\,$\mu$m bands are available from the more recent \textit{Spitzer} Extragalactic Representative Volume Survey \cite[SERVS,][]{Mauduit2012}. IRAC data are taken from the band-merged catalog of the XMM-LSS and ELAIS-N1 fields. For these merged catalogs we estimated the 90\% completeness magnitude limit values of 20.7, 20.8, 19.6, and 19.5\,mag for 3.6, 4.5, 5.8, and 8\,$\mu$m bands, respectively. The details of catalog construction are discussed in \S\,\ref{sec:clean_cat}.

\subsection{Spectroscopic redshifts}
The HSC-SSP DR1 catalog already includes the significant amount of spectroscopic data publicly available in these fields. For the ELAIS-N1 field data come from SDSS DR12 ($\sim$\,3100 spectroscopic sources). For the XMM-LSS field, data from various surveys are available: SDSS DR12, PRIMUS DR1 \citep{Coil2011}, VIPERS PDR1 \citep{Garilli2014}, and VVDS \citep{LeFevre2013}. The overlapping regions of these spectroscopic surveys with the HSC-SSP catalog of the XMM-LSS field contain a total of 49930 spectroscopic sources. We use these data to verify our cluster and photometric redshifts.

\subsection{Photometric redshifts}
Photometric redshifts are taken from the HSC-SSP DR1. The release contains different photometric redshifts computed using six independent codes. We use the values obtained through the DEmP algorithm \citep{Hsieh2014}. This code combines the nearest neighbor technique in multiple color-magnitude spaces with a polynomial fitting technique. The redshift value for each object is estimated using the 40 nearest neighbors in the nine-dimensional space (five magnitudes axes and $g-r$, $r-i$, $i-z^{\prime}$, $z^{\prime}-Y$ color axes using a PSF-matched aperture photometry) with a linear function. The redshifts obtained with this methodology show a good correlation with their corresponding spectroscopic values ($\sigma_{(1+z)}$\,=\,0.049), especially within our range of interest, between 0.2 and 1.1. Further details on the photometric redshift computation of HSC-SSP data are described in \cite{Tanaka2017temp}.

\subsection{Selection of galaxies with clean photometry}
\label{sec:clean_cat}
For selecting sources with reliable photometry, we follow the HSC SSP team recommendations. We select primary objects, which means objects that have been deblended and are in the inner patch and tract\footnote{Tracts are areas of 1.7\,$\times$\,1.7 square degrees of the sky predefined as an isolatitude tessellation. Those are further divided into 9\,$\times$\,9
sub-areas, each of which is 4200 pixels on a side, called patches.}. We also apply a set of flags to make sure that the objects do not suffer from problematic pixels (\texttt{flags\_pixel\_interpolated\_center}, \texttt{flags\_\-pixel\-\_\-edge}, \texttt{flags\_pixel\_cr\_center}), their centroids are correct (\texttt{centroid\_sdss\_flags}), their photometric measurements are good (\texttt{parent\_flux\-\_convolved\_flags}, \texttt{cmodel\_\-flux\-\_\-flags}), and they survived the junk suppression step\footnote{Subtraction of the very local sky to detect spurious sources. About 7\% of the objects are flagged as junk.} (\texttt{detected\_\-notjunk}) for the five photometric bands.

\begin{figure*}
\centering
\includegraphics[width=\hsize]{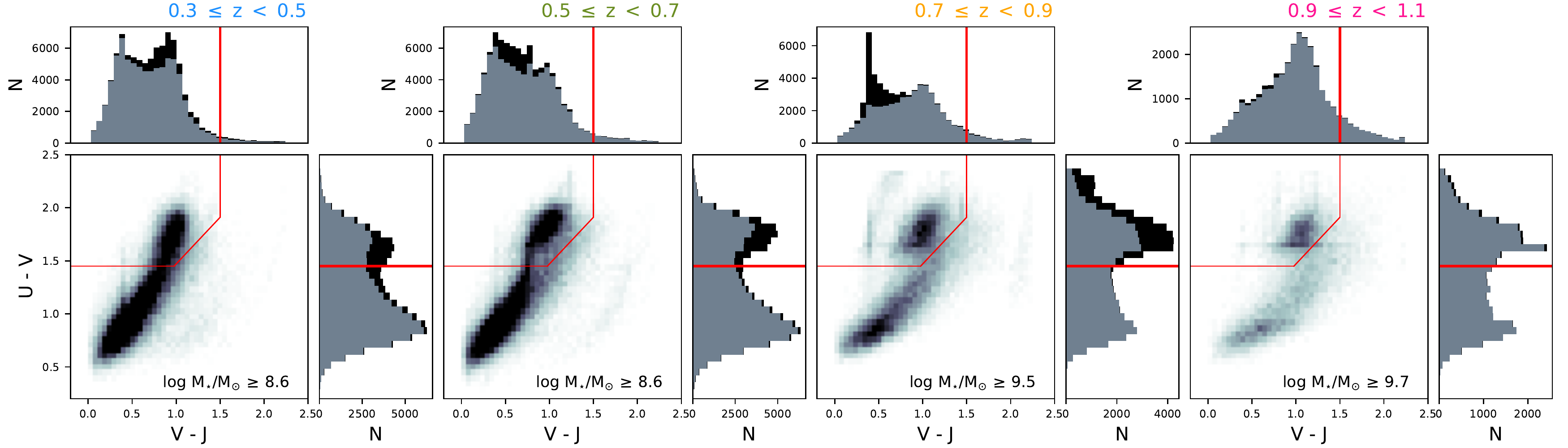}
\caption{UVJ color-color diagrams for different cluster redshift bins. Red lines separate UVJ-quiescent (top left region) and UVJ-star-forming (bottom and right regions) galaxies. The number in the bottom right corner of each central panel indicates the stellar mass completeness limit used in each redshift bin. Top and right panels show $V-J$ and $U-V$ histograms, respectively. Gray histograms show the final sample of galaxies employed, while black histogram shows the sample before removing the point-like sources (see \S\,\ref{sec:clean_cat}).}
\label{fig:uvj_colors}
\end{figure*}

The star/galaxy separation was performed using the \texttt{classification\_extendedness} parameter, which estimates the difference between photometry from PSF and CModel\footnote{Composite model photometry that fits a linear combination of exponential and de Vaucouleurs profiles convolved with PSF to objects \citep{Lupton2001,Abazajian2004,Bosch2017temp}.}. For compact sources, the CModel measurement approaches PSF photometry asymptotically and the difference in magnitude becomes small. In the HSC catalog, the extendedness parameter takes the value 0 when the difference is sufficiently small, and 1 otherwise. Thus, we remove from our sample only those objects which are classified as point sources in all five HSC bands. This stringent criterion for classifying an object as a star is due to faint small galaxies that are often classified as point sources. Furthermore, at faint magnitudes, there are many more galaxies than faint stars in high Galactic latitude fields and the overwhelming number of objects are galaxies. The number of stars rejected this way represent 6.5\% of our galaxy catalog. In Figure \ref{fig:uvj_colors} black and gray histograms show the distributions before and after removing the stars, respectively. In particular, the peak seen in the 0.7\,$<$\,$z$\,$<$\,0.9 panel, placed at rare blue $V-J$ colors and red $U-V$ colors, shows the importance of removing such point sources from our catalog. Taking advantage of the ancillary J and K data, we perform a sanity check of the star classification based on the extendedness parameter. We find that 90, 77, 87, and 59\%\footnote{The total number of sources classified as stars are 7135, 4077, 4215, and 550 at z\,$\sim$\,0.4, 0.6, 0.8, and 1, respectively. These numbers show that HSC photometric redshift estimate for stars tend to put  them preferably at z\,$\sim$\,0.4 and 0.8.} of the objects classified as stars at z\,$\sim$\,0.4, 0.6, 0.8, and 1, respectively, show blue $J-K$ colors ($J-K$\,$<$\,1).

Once we have our clean optical photometric catalog ($griz^{\prime}Y$), we cross-match it with the infrared catalogs. We use the nearest neighbor technique using a maximum separation radius of 1$''$ for UKIDSS, and 1.5$''$ for IRAC/SWIRE+SERVS. The absolute separation distribution peaks at 0.06$''$ and at 0.2$''$ for UKIDSS and IRAC/SWIRE+SERVS, respectively.

\begin{figure*}
\centering
\includegraphics[width=\hsize]{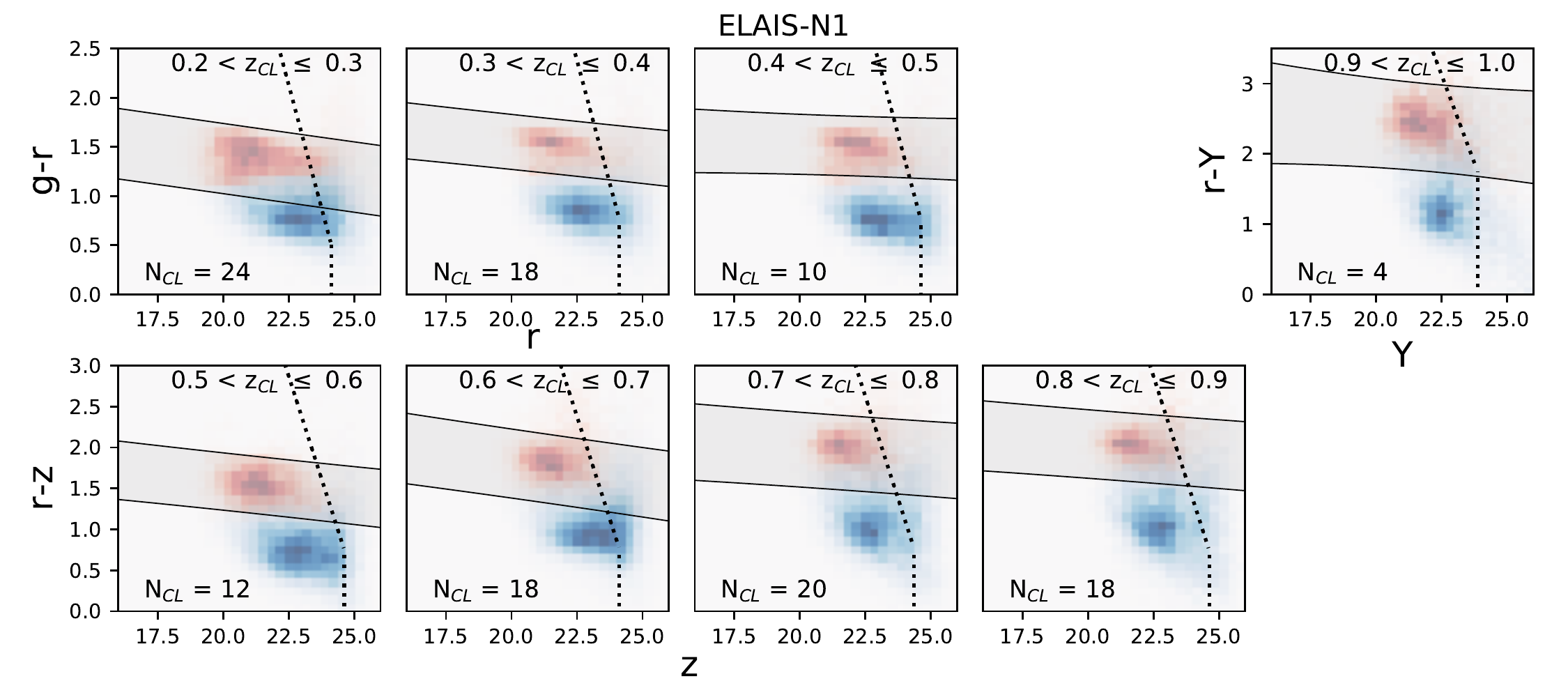}
\includegraphics[width=\hsize]{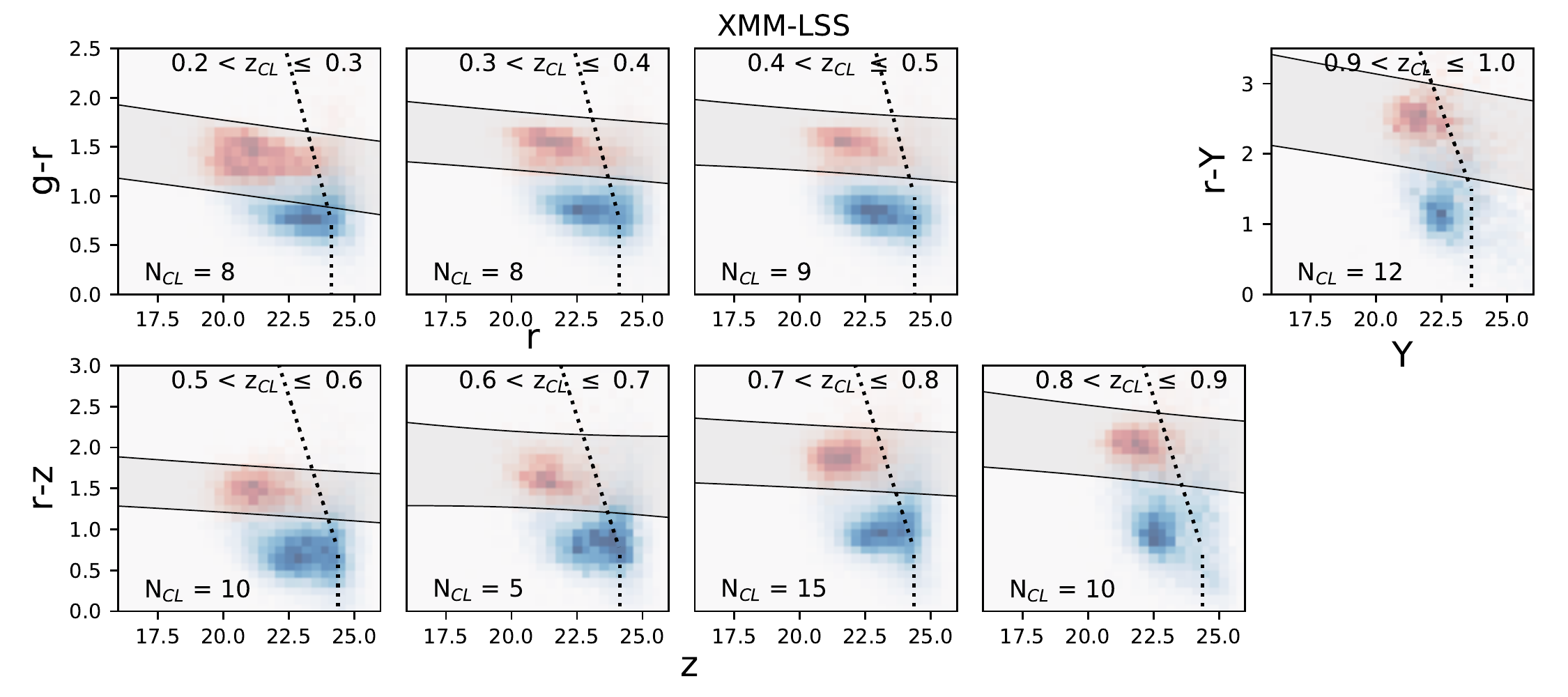}
\caption{Color-magnitude diagrams (CMD) for different cluster redshift bins in the two fields: ELAIS-N1 (the two top rows) and XMM-LSS (the two bottom rows). Red and blue colors represent the 2D histograms of UVJ-quiescent and UVJ-star-forming galaxies, respectively. The number in the bottom left corner of each panel indicates the number of clusters in each bin, which are stacked and used to calculate the red sequence. The shaded region between the two black curves defines the red-sequence region, i.e., the 98\% confidence interval of the fit (used to determine the final cluster redshift, see points 1 and 4 in \S\ref{sec:cl_sample}). The dotted lines indicate the magnitude completeness limit.}
\label{fig:cmd_rs_uvj}
\end{figure*}


\begin{figure}
\centering
\includegraphics[width=\hsize]{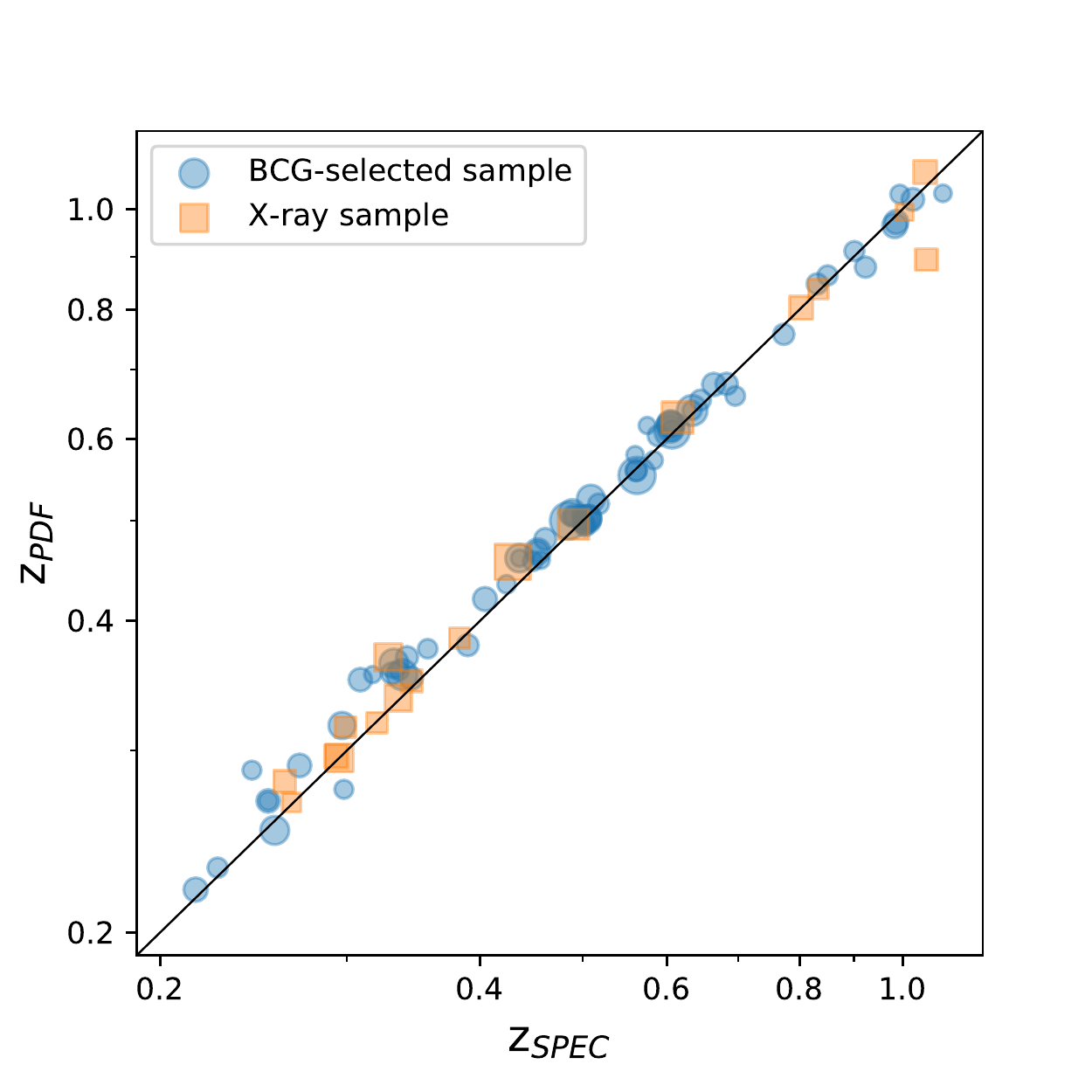}
\caption{The photometric cluster redshift, estimated as detailed in Section \ref{sec:cl_sample}, versus the spectroscopic cluster redshift obtained from a subsample of spectroscopic galaxies within 15000\,km\,s$^{-1}$ of BCGs that have spectroscopic redshifts (blue circles) and from XMM-LSS survey (orange squares, \cite{Adami2011,Willis2013} ). The size of the symbols is scaled with the cluster richness. The total number of clusters shown is 80, as we do not plot the 11 clusters from the BCG-selected sample that are already included in the X-ray sample.}
\label{fig:zspec_vs_zphot}
\end{figure}


\subsection{Cluster sample and membership}
\label{sec:cl_sample}

The clusters analyzed in this study are from two of the six SpARCS \citep{Muzzin2009,Wilson2009} fields. Clusters are originally found using a slightly modified version of the cluster red-sequence algorithm developed by \cite{Gladders2000,Gladders2005}. The code is described in more detail in \cite{Muzzin2008}. The only modification of the algorithm from \cite{Muzzin2008} is that of the optical band in the color, $z^{\prime}$\,-\,3.6$\mu$m instead of $R$\,-\,3.6$\mu$m. From the full sample of cluster candidates we choose those with richness $N_{\mathrm{red}}$\,$\footnote{Number of close neighbors of a red galaxy.}>$\,6.0 (equivalent to $B_{\mathrm{gc}}\footnote{Cluster-center galaxy correlation amplitude, usually used to quantify the environment of radio galaxies.}\,\gtrsim\,330$), in order to reduce the number of false clusters. The approximate cluster mass limit is 9\,$\times$\,10$^{13}$\,M$_{\odot}$, computed using Equation 9 from \cite{Muzzin2007} which relates $B_{\mathrm{gc}}$ and $M_{200}$ (the mass within a radius that encloses a density 200 times the critical density of the Universe). 

The original SpARCS sample has 188 and 167 cluster candidates in the ELAIS-N1 and XMM-LSS fields up to redshift 1.2 and $N_{\mathrm{red}}>$\,6.0, respectively. This is chosen as our parent sample, which we further refine in several ways. The first is that the redshifts were defined using only the red-sequence, but now we have full photometric redshift information. We use this information to refine the cluster redshifts as follows:

1. We compute stacked red-sequences for clusters in redshift bins of 0.1 from $z_{\mathrm{SpARCS}}$\,=\,0.3 up to 1.1, where $z_{\mathrm{SpARCS}}$ is the original SpARCS catalog cluster photometric redshift, by considering galaxies at a maximum distance from the cluster center of $R_{\mathrm{MAX}}$\,=\,1.5\,Mpc and a redshift range defined by $\sigma_{\Delta z}$\,=\,0.05, where we use both photometric and spectroscopic redshifts. Note that we use observed magnitudes, so the color-magnitude diagram employed is different for different redshift bins. We fit a line to the galaxies within the red color region (1.1-1.8, 1.3-1.9, and 1.3-1.9 in $g-r$ vs. $r$ for $z_{\mathrm{SpARCS}}$\,$\sim$\,0.35, 0.45, and 0.55, respectively; 1.1-1.9, 1.4-2.1, 1.6-2.3, and 1.7-2.4 in $r-z^{\prime}$ vs. $z^{\prime}$ for $z_{\mathrm{SpARCS}}$\,$\sim$\,0.65, 0.75, 0.85, 0.95, respectively; and, 1.9-3.0 in $r-Y$ vs. $Y$ for $z_{\mathrm{SpARCS}}$\,$\sim$\,1.05) and define the 98\% confidence intervals of the fit as the red-sequence region.

2. For each cluster, we utilize galaxies interior to the maximum cluster-centric radius of $R_{\mathrm{MAX}}$\,$=$\,0.5\,Mpc and $\sigma_{\Delta z}$\,=\,0.2 within the fitted red-sequence region to define an initial cluster redshift, $z_{\mathrm{CL}}$, which is the peak of the redshift probability density function resulting from the sum of the $z_\mathrm{{PDF}}$ of all these red-sequence galaxies. Here, we reduce the distance to the cluster center to avoid contamination due to nearby clusters in the projected plane, and use a wider redshift range to take into account the uncertainty in the original SpARCS redshift estimation.

3. For clusters with catastrophic photometric redshifts, defined as either a difference between the redshift of the SpARCS BCG and $z_{\mathrm{SpARCS}}$ of larger than 0.15\,$\times$\,(1+$z$), or when less than three red-sequence galaxies are found in the previous step, we define a provisional $z_{\mathrm{CL}}$ as the peak of the redshift PDF resulting from the sum of the $z_{\mathrm{PDF}}$ of all galaxies, not only the red-sequence ones. We perform step 2 again changing $z_{\mathrm{SpARCS}}$ to the provisional $z_{\mathrm{CL}}$, and estimate the initial $z_{\mathrm{CL}}$ when more than two red-sequence galaxies are found. 

4. We repeat steps 1 and 2 changing $z_{\mathrm{SpARCS}}$ to the initial $z_{\mathrm{CL}}$ to determine the final $z_{\mathrm{CL}}$. The red-sequence region obtained after the second iteration is shown in Figure \ref{fig:cmd_rs_uvj} as the gray shaded area between the black curves. From the original SpARCS catalog, the persistent catastrophic $z_{\mathrm{SpARCS}}$ clusters (i.e., when less than three red-sequence galaxies are found after the second iteration) are flagged with value 9, and those clusters outside the HSC-SSP field of view are flagged with value -99. Clusters flagged with either of these values are not included in the analysis of this paper. 

After these cuts, the final sample of clusters with which we work comprises of 127 in ELAIS-N1 and 82 in XMM-LSS. We then select potential cluster galaxies by selecting galaxies that are within the cluster photometric redshift slices.  
We adopt spectroscopic redshifts when available and DEmP photometric values for the rest of the galaxies. The final galaxy catalog includes a total of 1264407 galaxies, with 60\% of them in the ELAIS-N1 field and the remainder in the XMM-LSS field. 
For each cluster, we use galaxies that are included in a slice defined by the $z_{\mathrm{CL}}$\,$\pm$\,0.05\,$\times$\,(1+$z_{\mathrm{CL}}$) range and up to 20\,Mpc in cluster-centric radius. 
We note that since the slices associated with the clusters can overlap, a galaxy in overlapping areas can be included in multiple slices.

We compare our photometrically determined cluster $z_{\mathrm{CL}}$ with available spectroscopic measurements from the XMM-LSS survey from \cite{Adami2011} and \cite{Willis2013} for 20 clusters, and we obtain a good correlation with a slope of 1.04 and RMS\,=\,0.015. Moreover, we take advantage of the large spectroscopic sample to double check our $z_{\mathrm{CL}}$ values. We identify clusters whose BCGs have spectroscopic redshifts and compute the peak of the distribution of the spectroscopic redshifts, including that of the BCG, within $R_{200}$ and $\sigma_{z}$\,=\,0.05. Considering the 70 BCG-selected clusters with three or more spectroscopic redshifts, we find that RMS of the $z_{\mathrm{CL}}$ versus $z_{\mathrm{SPEC}}$ relation is 0.02. We note that 11 out of these 70 clusters overlap with the XMM-LSS cluster sample. These comparisons show that the photometric cluster redshift estimates are very accurate.

\section{Fraction of star-forming galaxies}

\subsection{Stellar masses and UVJ rest-frame colors}
\label{sec:mass_uvj}
With our 11-band ($grizYJK,$\,3.6,\,4.5,\,5.8,\,8.0\,$\mu$m\footnote{Note that $\sim$\,10\% of clusters have only 9 bands as there is a non-negligible area in the XMM-LSS field without J and K data. We compare the stellar masses estimated with and without J and K bands, for the galaxies with these bands available, and we obtain a good correlation with a slope of 0.97 and RMS\,=\,0.3}) catalog we estimate stellar masses and rest-frame colors. For both computations, we use SED-fitting software and the photometric/spectroscopic redshifts of the galaxies themselves. 

To determine stellar masses we use the \textit{LePhare} code \citep{Arnouts1999,Ilbert2006} to fit the stellar part of the spectrum with \cite{Bruzual2003} population synthesis models, with star formation histories exponentially declining as $e^{-t/\tau}$. The complete template library was built considering solar metallicity, nine different values of $\tau$ (from 0.1 to 30\,Gyr) with 57 steps in age, and the extinction law of \cite{Calzetti2000} with values of $E(B-V)$ ranging from 0 to 1. The templates also account for the contribution of the emission lines to the flux. We assume the initial mass function (IMF) of \cite{Chabrier2003}. 

The rest-frame $U-V$ and $V-J$ colors are computed with the EAZY \citep{Brammer2008} program. We use a combination of a reduced set of seven templates, similar to the \cite{Brammer2008} original set: five are the reduced set from \cite{Grazian2006} templates but with emission lines to match \cite{Ilbert2009}'s prescription, the sixth template accounts for dusty galaxies that do not appear in semi-analytic models, and the final template is an evolved \cite{Maraston2005} model to be red enough for massive old galaxies at $z$\,$<$\,1. Our lower redshift limit of $z$\,$=$\,0.3 is set to ensure a good estimation of rest-frame magnitudes, as the bluest filter we have available is the $g$ band. 

These rest-frame colors allow us to classify galaxies as UVJ-star-forming and UVJ-quiescent, as shown by the UVJ color-color diagrams of Figure \ref{fig:uvj_colors} and the 2D colored distributions of Figure \ref{fig:cmd_rs_uvj}. The cuts used to define the star-forming and quiescent regions are slightly modified from the calibration made by \cite{Williams2009} to fit the bimodality in our data. The criteria are $U-V$\,$>$\,1.45, $V-J$\,$<$\,1.5, and $U-V$\,$>$\,0.88\,$\times$\,$(V-J)$\,$+$\,0.59. Figure \ref{fig:cmd_rs_uvj} shows a good agreement between the classification made using UVJ colors and the red sequence computed in Section\,\ref{sec:cl_sample} for the selection of cluster galaxies, where most of the quiescent population fall within the red sequence region and some of the SF galaxies populate its faint end. 

\subsection{Stellar mass completeness}
A careful study of the evolution of the SF fraction needs to take into account different completeness limits at different redshifts. Therefore, we have examined the stellar mass differential distributions and define the stellar mass threshold for each redshift range as the peak and, when studying evolution, compare samples in equivalent stellar mass bins. The values obtained for the stellar mass limit are log\,$M_{\star}$/M$_{\odot}$\,=\,8.6, 8.6, 9.5, and 9.7 for $z$\,$\sim$\,0.4, 0.6, 0.8, and 1, respectively.

\subsection{Background subtraction}

Our estimation of the number of star-forming galaxies over the total number of cluster galaxies is contaminated by the inclusion of field galaxies due to our selection method based on photometric redshifts. We make use of the large coverage of our catalog to correct for this background contamination. We generate background field control samples for each cluster considering those galaxies in an annular area of inner $R/R_{200}$ radius 6 and outer $R/R_{200}$ radius 9. The selection of these radii is a compromise between being sufficiently large that no cluster galaxies are included and not so large that the correction for the area outside our spatial coverage and the contamination due to nearby clusters are small. The results we show are stable against changes in the size of the area used for defining the background control sample, which we expect, given the surface galaxy density as a function of the radius is almost flat beyond $\sim$\,2.5\,Mpc. The subtraction of the background signal is performed when computing the fraction of SF galaxies, as described in \S \ref{sec:fsf_bayes}.

\subsection{The SF fraction with Bayesian inference}
\label{sec:fsf_bayes}
The fraction of SF galaxies is, by definition, in the $[\,0, 1]$ range, but the presence of background sources could yield unphysical values of negative fractions or values $>$\,1 (a subsample is larger than the whole sample). This occurs when the following formula
\begin{equation}
\label{eq:f_sf}
f_{SF}(Cluster)\,=\,\dfrac{n_{SF}(Cluster+Back)\,-\,n_{SF}(Back)}{n_{TOT}(Cluster+Back)\,-\,n_{TOT}(Back)}
\end{equation}
is used to compute the SF fraction in clusters. To avoid nonsense fractions, we use Bayesian inference to calculate the posterior probability that clusters have a fraction of SF galaxies $f_{SF}$, given the same quantities that appear in Equation\,\ref{eq:f_sf}. We followed the methodology described in \cite{Andreon2006} and the mathematical details expounded in \cite{DAgostini2004}. However, instead of taken the median value from the posterior distribution as in \cite{Andreon2006}, our estimate of the cluster $f_{SF}$ is the peak of the posterior distribution and its uncertainty is defined as the FWHM interval.

\begin{figure*}
\centering
\includegraphics[width=\hsize]{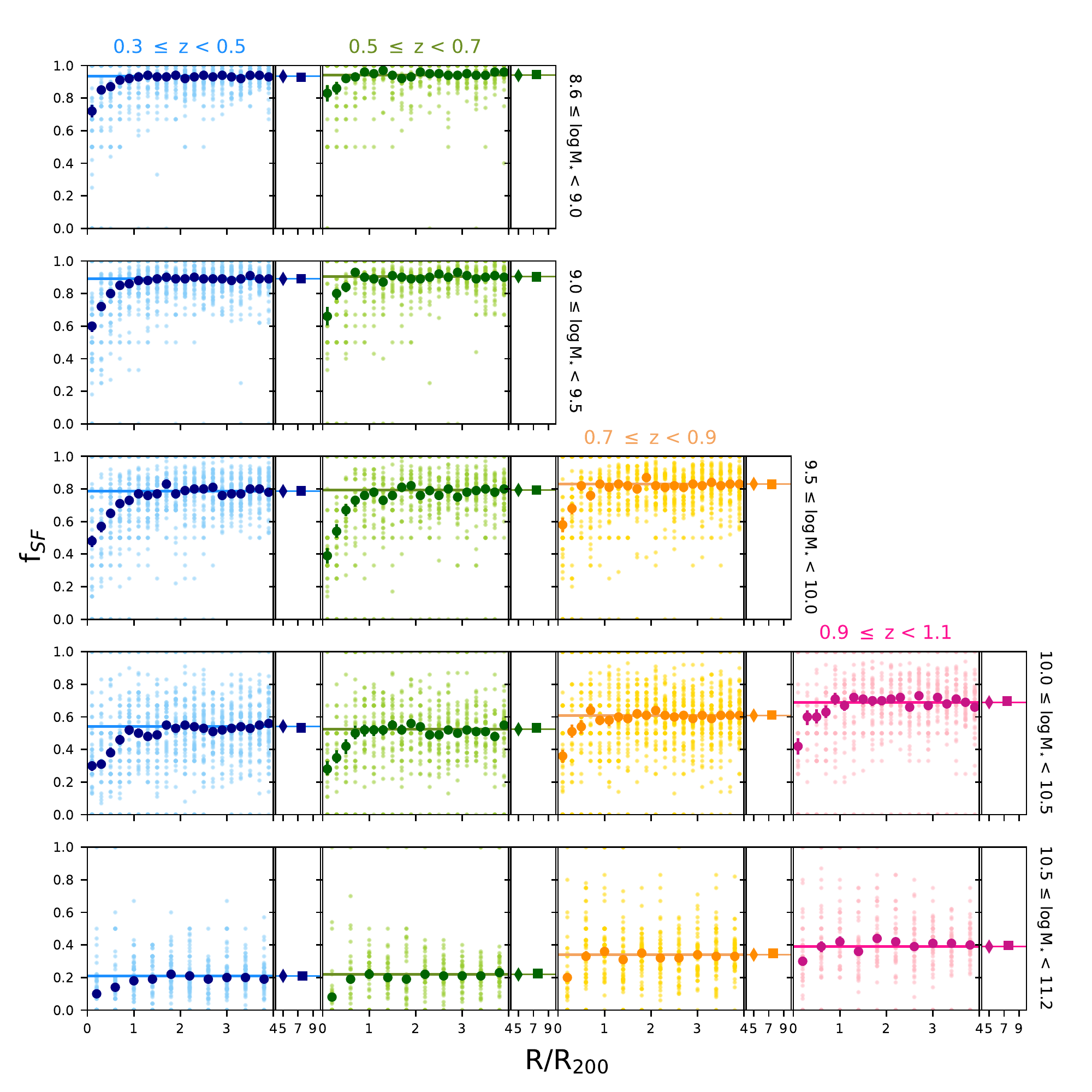}
\caption{Fraction of SF galaxies as a function of the distance to the center of the cluster ($R/R_{200}$). Panels from left to right increase in redshift and from top to bottom in stellar mass, as indicated by top titles and labels on the right. Star-forming galaxies are selected by their UVJ rest-frame colors. As described in \S\, \ref{sec:fsf_bayes}, 
the fraction is estimated for individual clusters (smaller light color circles) using Bayesian inference to account for the background contamination. The composite fraction is computed by stacking galaxies from all clusters and applying the same Bayesian formula (bigger dark circles), and the errors are the FWHM of the posterior distribution.  
In the right secondary panels, diamond and square symbols indicate the mean $f_{SF}$ of the ``field environment'' (also the horizontal line in main panels), defined by the region 4\,$\leq$\,$R/R_{200}$\,$<$\,6, and the control ``background field'', defined by the region 6\,$\leq$\,$R/R_{200}$\,$<$\,9, respectively. The width of $R/R_{200}$ bins is 0.2, except for the highest stellar mass bin where is increased to 0.4.}
\label{fig:fuvj_vs_rr200}
\end{figure*}

We compute the $f_{SF}$ for each cluster as a function of the cluster-centric distance in four cluster redshift bins (0.3\,$\leq$\,$z$\,$<$\,0.5, 0.5\,$\leq$\,$z$\,$<$\,0.7, 0.7\,$\leq$\,$z$\,$<$\,0.9, 0.9\,$\leq$\,$z$\,$<$\,1.1) and five stellar mass bins (8.6\,$\leq$\,log\,$M_{\star}$/M$_{\odot}$\,$<$\,9.0, 9.0\,$\leq$\,log\,$M_{\star}$/M$_{\odot}$\,$<$\,9.5, 9.5\,$\leq$\,log\,$M_{\star}$/M$_{\odot}$\,$<$\,10.0, 10.0 $\leq$\,log\,$M_{\star}$/M$_{\odot}$\,$<$\,10.5, 10.5\,$\leq$ log\,$M_{\star}$/M$_{\odot}$\,$<$\,11.2). The $R/R_{200}$ bin width is 0.2, except for the highest stellar mass bin where the width is increased to 0.4. $R_{200}$ is computed using Equation 2 from \cite{Muzzin2008} which relates $B_{\mathrm{gc}}$ and $R_{200}$. The results are presented in Figure \ref{fig:fuvj_vs_rr200}, where the $f_{SF}$ values for individual clusters are shown as light colored points. To estimate the composite $f_{SF}$ in each radial bin, we stack the galaxies from all clusters in each bin and apply the same Bayesian inference approach as for individual clusters. The result obtained for the composite $f_{SF}$ is completely compatible with the mean of the $f_{SF}$ of individual clusters.
The composite values are plotted as dark circles. 
The diamond symbols and the light horizontal lines represent the near-field, from $R/R_{200}$\,=\,4 to 6, for which we use as a comparison ``field environment'', and refer to as ``field'' hereafter. These $f_{SF}$ field values have also been background corrected. The squares denote the value of the ``background field'' region employed for the background correction from $R/R_{200}$\,=\,6 to 9.
We note here that our $f_{SF}$ composite values at $R/R_{200}$\,$\gtrsim$\,2 lie along the value of the field. We note that this behavior is not a bias of the method because the points have net counts over that of the background field values.

\subsection{Elimination of BCGs}

As we are analyzing the effect of the cluster environment on the quenching of the star formation in its member galaxies, we note that the brightest cluster galaxies (BCGs) are special objects whose formation and evolution is complex: they are involved in a combination of galaxy merging, cluster cooling flows, AGN feedback and star formation processes \cite[e.g.,][]{Rawle2012,Webb2015,McDonald2016}. Thus, these processes will affect our results, particularly in a sensitive region as is the innermost $R/R_{200}$ of the highest stellar mass bin. We select BCGs as the brightest galaxy in the same magnitude used for estimating the red sequence (i.e., $r$, $z'$, and $Y$ bands for $z$\,=0.3-0.6,0.6-1.0, and 1.0-1.1, respectively) and remove them from our cluster galaxy sample.

\section{Results}

In this section we present the results on the dependence of the SF fraction on galaxy stellar mass, distance to the cluster center, and redshift. Our aim is to trace how the evolution of the fraction of galaxies that quench their star formation is connected to its cluster environment by using cluster-centric radius. For this purpose, we need to distinguish the effect of the cluster environment from the secular evolution of galaxies. Our approach is to measure the SF fraction as a function of the environment at fixed stellar mass and vice versa.

\begin{figure*}
\centering
\includegraphics[width=\hsize]{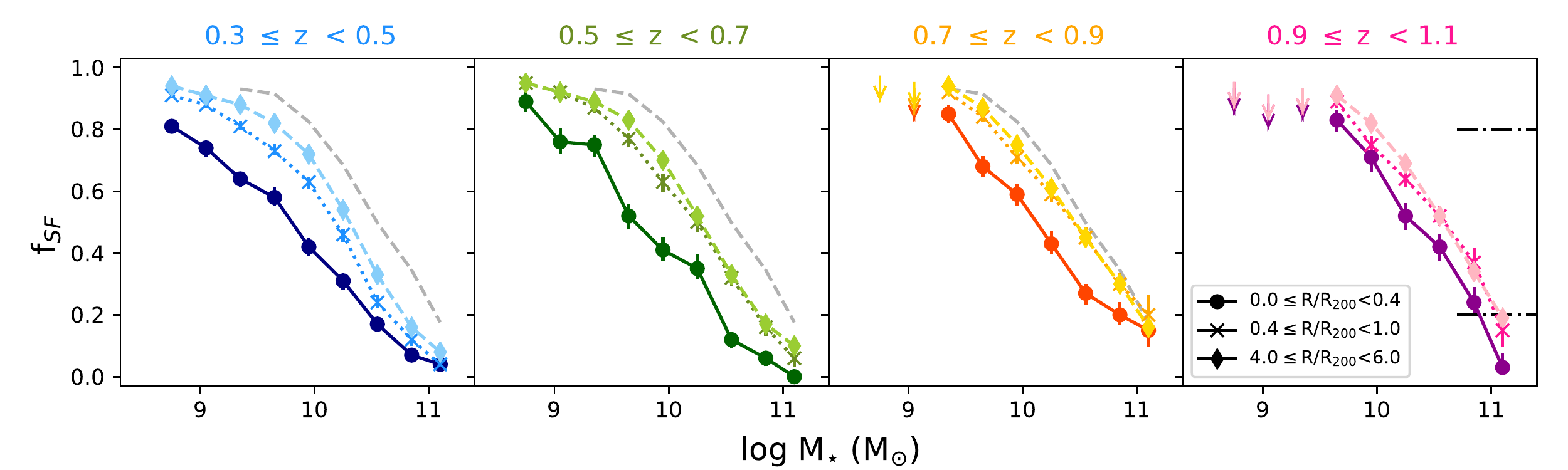}
\caption{Fraction of SF galaxies as a function of stellar mass in three different environmental regions: 0\,$\leq$\,$R/R_{200}$\,$\leq$\,0.4 (\textit{cluster core}, dark colored circles), 0.4\,$\leq$\,$R/R_{200}$\,$\leq$\,1 (\textit{cluster outskirts}, medium colored crosses), and 4\,$\leq$\,$R/R_{200}$\,$\leq$\,6 (\textit{field}, light colored diamonds). We include the field value at the highest redshift bin as a reference in the other panels (dashed gray curve). Arrows indicate the upper limit of the SF fraction at stellar mass bins where the star forming population is complete, but the passive population is not. Panels from left to right increase in redshift as indicated by top titles. The dot-dashed horizontal lines in the right panel are the locus of the fractions used to define $M^{*}_{\mathrm{INI}}$ and $M^{*}_{\mathrm{END}}$. As in the previous figure, the fraction is computed summing up galaxies from every cluster and using Bayesian inference to account for the background contamination, and the errors are the FWHM of the posterior distribution. The width of stellar mass bins is 0.3\,dex.}
\label{fig:fuvj_vs_m}
\end{figure*}

\subsection{The fraction of star-forming galaxies: environment and mass dependency}
\label{sec:fsf_m_and_env}

The large amount of data available allow us to study the $f_{SF}$ in four epochs, in $R/R_{200}$ bins of 0.2 from the cluster core to the field, and over a large range of galaxy stellar masses at the same time, while maintaining small errors, as shown in Figure \ref{fig:fuvj_vs_rr200}. Note that for each redshift we only show stellar mass bins down to their completeness limit. The panels in this figure give us an idea of the role of the stellar mass and the environment in quenching star formation activity and their evolution. 

Focusing on a single epoch, i.e., looking at one column in Figure \ref{fig:fuvj_vs_rr200}, the composite $f_{SF}$ decreases drastically from 1 to below 0.5 from low to high stellar masses. This drop is more dramatic for the lower redshift bins: from log\,$M_{\star}$/M$_{\odot}$\,$\simeq$\,9.7 to 10.8 the field SF population is reduced by a factor of $\sim$\,4. Over the redshift range of 0.7 to 0.9 the field $f_{SF}$ is reduced by a factor of $\sim$\,3 over the same mass range. The decline of the $f_{SF}$ from low to high stellar masses is evident not only in the reference value of the field, but also in the cluster core, where the SF population decreases by a factor of $\sim$\,4 from log\,$M_{\star}$/M$_{\odot}$\,$\simeq$\,9.7 to 10.8 in all three lower redshift bins. This trend is seen more clearly in Figure \ref{fig:fuvj_vs_m}, where the $R/R_{200}$ dimension has been binned into three regions and the $f_{SF}$ is presented as a function of the stellar mass using narrower stellar mass bins. The regions used are: the cluster core up to $R/R_{200}$\,$=$\,0.4; the cluster outskirts combining 0.4\,$\leq$\,$R/R_{200}$\,$<$\,1; and the field as estimated between 4\,$\leq$\,$R/R_{200}$\,$<$\,6. We deliberately exclude the plot of the 1\,$\leq$\,$R/R_{200}$\,$<$\,4 region because it overlaps with the field curve, as one can infer from Figure \ref{fig:fuvj_vs_rr200}. 

\begin{figure}
\centering
\includegraphics[width=0.9\hsize]{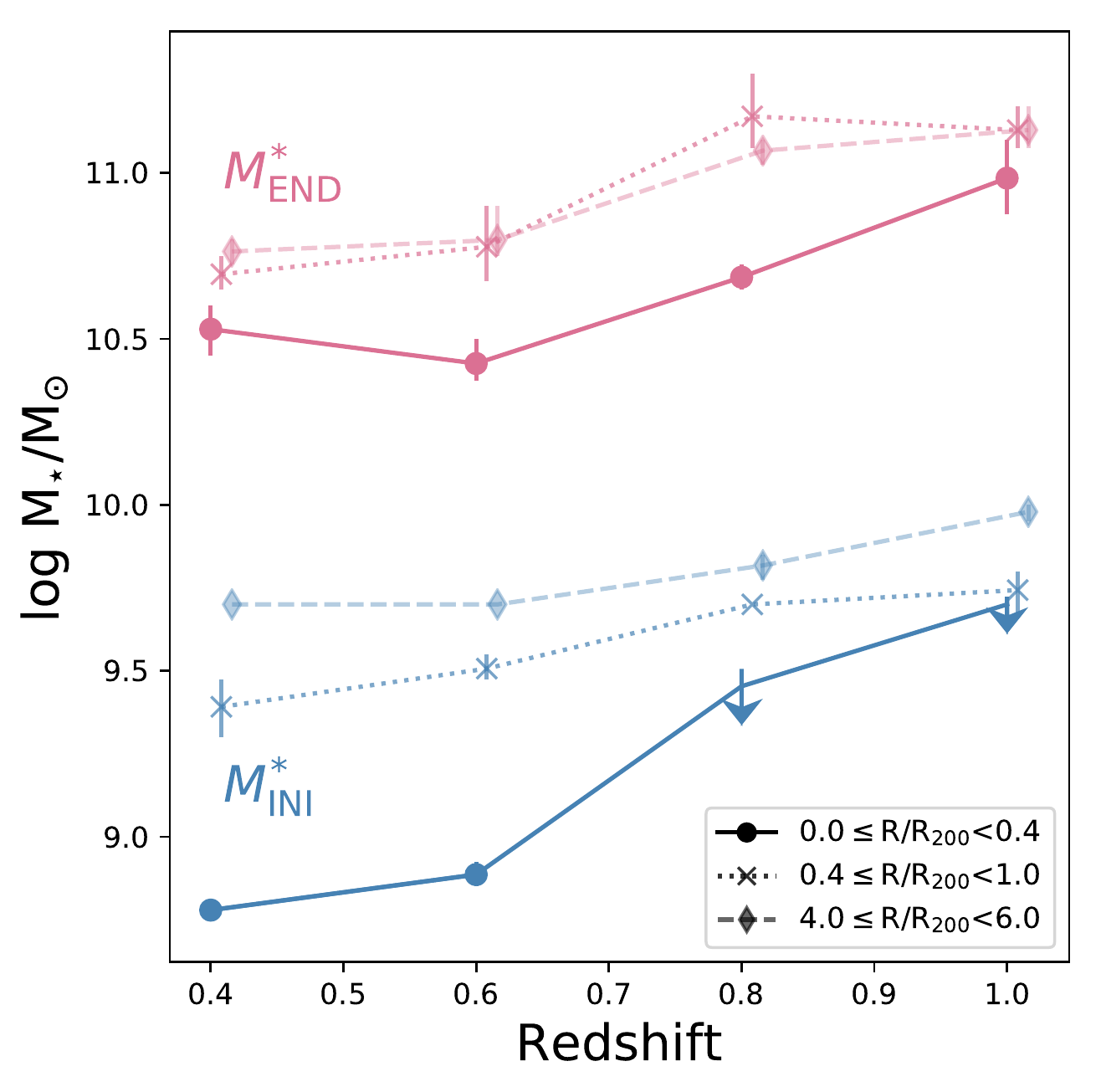}
\caption{$M^{*}_{\mathrm{INI}}$ (blue bottom lines) and $M^{*}_{\mathrm{END}}$ (pink top lines) values, as defined in \S\,\ref{sec:fsf_m_and_env}, as a function of redshift for three different environments: 0\,$\leq$\,$R/R_{200}$\,$\leq$\,0.4 (\textit{cluster core}, dark colored circles), 0.4\,$\leq$\,$R/R_{200}$\,$\leq$\,1 (\textit{cluster outskirts}, medium colored crosses), and 4\,$\leq$\,$R/R_{200}$\,$\leq$\,6 (\textit{field}, light colored diamonds). The two arrows indicate that the measured value is just below our mass completeness limit.}
\label{fig:mini_mend_vs_z}
\end{figure}

With these data, we can quantify two characteristic stellar masses: the mass at which the quenching process starts, $M^{*}_{\mathrm{INI}}$, and the mass at which almost all galaxies are quenched, $M^{*}_{\mathrm{END}}$, using a simple procedure. We estimate $M^{*}_{\mathrm{INI}}$ at where $f_{SF}$ becomes lower than 0.8, and $M^{*}_{\mathrm{END}}$ at where $f_{SF}$ goes below 0.2 (as marked in the right panel of Figure \ref{fig:fuvj_vs_m}). The result is presented in Figure \ref{fig:mini_mend_vs_z}. Note that all measured values are above our stellar mass completeness limit. Within our redshift range of study, we observe a clear trend in both measures: these characteristic stellar masses evolve to lower values towards lower redshifts.

Figure \ref{fig:fuvj_vs_m} not only reveals information on the quenching effects caused by stellar mass, but also about the environmental quenching process that takes place in the cluster environment. In these four panels, the curves representing the cluster core show a shortage of the SF population compared to the field at all stellar masses. 
We observe little difference in the $f_{SF}$ vs. stellar mass relation between the outskirt of the cluster and the field in Figure \ref{fig:fuvj_vs_m} for the two higher redshift bins.
However, the relation for the cluster outskirts becomes slightly lower for the 0.5\,$\leq\,z\,<$\,0.7 bin, and significantly lower for the lowest redshift bin at all stellar masses. This indicates that little or no environmental quenching effect as measured by $f_{SF}$ is evident beyond $R/R_{200}$\,$\sim$\,1 for clusters at $z\,\geq$\,0.7. 



\begin{figure*}
\centering
\includegraphics[width=\hsize]{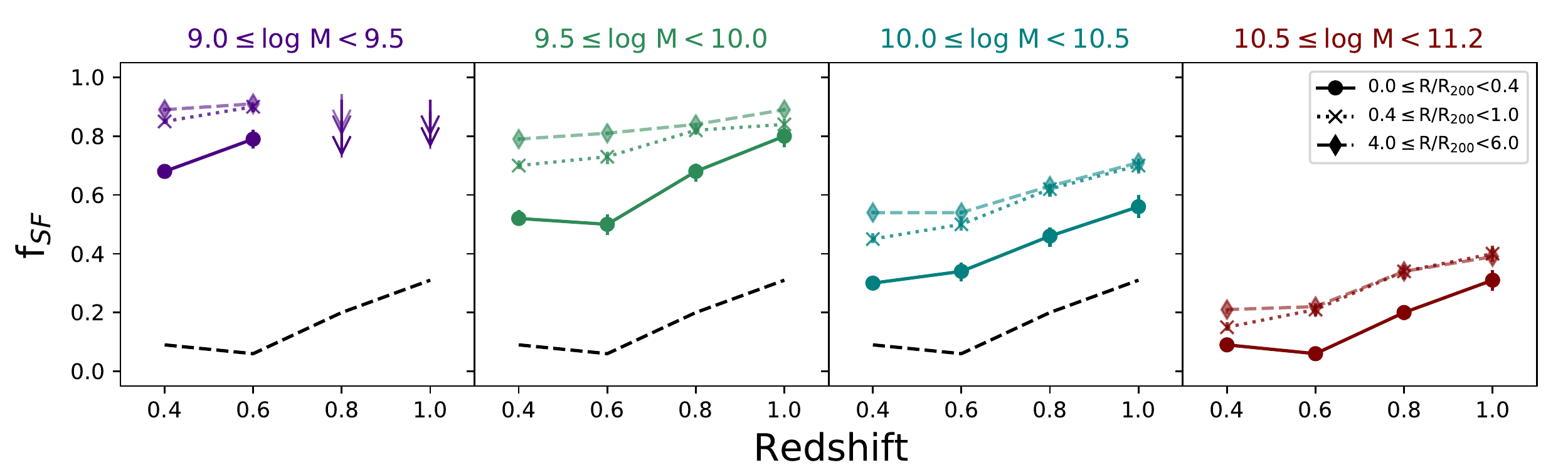}
\caption{Evolution of the fraction of SF galaxies in four stellar mass bins and in three different environments: 0\,$\leq\,R/R_{200}\,\leq$\,0.4 (\textit{cluster core}, dark colored circles), 0.4\,$\leq\,R/R_{200}\,\leq$\,1 (\textit{cluster outskirts}, medium colored crosses), and 4\,$\leq\,R/R_{200}\,\leq$\,6 (\textit{field}, light colored diamonds). Panels from left to right increase in stellar mass as indicated by top titles. We include the cluster value at the highest stellar mass bin as a reference in the other panels (dashed black curve). As in previous figures, the fraction is determined from the composite cluster galaxy sample and applying Bayesian inference. The errors are the FWHM of the posterior distribution.}
\label{fig:fuvj_vs_z}
\end{figure*}


\subsection{Evolution of the SF fraction: environment and mass dependency}
\label{sec:boeffect}

Focusing on a single stellar mass bin in Figure \ref{fig:fuvj_vs_rr200}, i.e., analyzing rows, we can study the evolution of the SF fraction with redshift. In Figure \ref{fig:fuvj_vs_z}, in a fashion similar to previously done in Figure \ref{fig:fuvj_vs_m}, we bin the $R/R_{200}$ dimension in three regions to show the fraction of SF galaxies as a function of epoch. 
In these panels we observe the strong evolution of the fraction of SF galaxies with redshift, first seen by \cite{Butcher1984}, and subsequently many other studies \citep[e.g.,][]{Balogh1999,Poggianti2006,Saintonge2008,Brodwin2013,Wagner2017,Jian2018}. Furthermore, our large sample of clusters covering a relatively large range of redshift demonstrates clearly the dependence of this evolution on both galaxy stellar mass and the cluster-centric radius. In the far right panel of Figure\,\ref{fig:fuvj_vs_z}, we see how more massive cluster galaxies (log\,$M_{\star}$/M$_{\odot}\,\geq\,$10.5) are mostly quenched by z\,$\sim$\,0.6 ($f_{SF}$\,$\leq$\,0.1). We find that the change of $f_{SF}$ is fractionally larger for higher stellar mass galaxies, in both the field and the cluster environment. 

The large area coverage of our data allows us to analyze large radii and define the field and the cluster in a consistent way, being able to confidently compare the $f_{SF}$ evolution curves of the different environments. The four panels in Figure \ref{fig:fuvj_vs_z} show how the cluster environmental impact shifts the $f_{SF}$ to lower values compared to the field for all redshift and stellar mass bins, but such a shift also depends on redshift and stellar mass. The shift between the $f_{SF}$ of the cluster core and the field is fractionally larger for lower redshift bins and higher mass galaxies. Thus, the environment causes a steepening of the $f_{SF}$ vs. redshift curves.


\subsection{Possible systematics and their effects on the $f_{SF}$ results}

\subsubsection{Photo-z uncertainties for different populations}
A quantity that can be different for SF and quiescent populations is the photometric redshift dispersion $\sigma_{\Delta z}$, as the form of the SED of quiescent galaxies makes it easier to estimate the correct redshift through photometric data fitting. We employ the available spectroscopic data to examine the dispersion of the $z_{\mathrm{SPEC}}$ vs. $z_{\mathrm{PHOT}}$ relation for these two populations. We find that, though the SF sample contains more outliers (20\% versus 14\% of the galaxies are outside the 0.05$\times$(1+z) interval), the $\sigma_{\Delta z}$ for the SF and quiescent populations are not significantly different and, down to our mass completeness limit, are less than 0.05. This means that we are not biasing our result when selecting cluster galaxies in redshift slices of $\sigma_{\Delta z}$\,=\,0.05. 

\subsubsection{Cluster sample: missing clusters?}
As mentioned in Section\,\ref{sec:cl_sample}, our initial sample of clusters is limited to those with $N_{red}>$\,6.0, and then it is cleaned by excluding clusters that fall outside of the HSC area or do not have enough red-sequence galaxies to determine $z_{CL}$ by means of a photometric redshift distribution. This procedure helps us to reduce the contamination of false clusters, but it might bring the question of the lack of clusters with very blue populations, as by construction our sample is biased towards clusters with well-defined red sequences. Nevertheless, the richness limit of $N_{red}=$\,6.0 is rather low and we require only a small number of red sequence galaxies to estimate the cluster redshift, just $>\,$3. Even for the case of a cluster with a very high fraction of blue galaxies, we still expect it to have a number of red galaxies, and, since we find no literature reporting the detection of such blue clusters we can safely argue that the number of missing clusters is very low. In addition, when comparing X-ray selected clusters from the XMM-LSS distant cluster survey and clusters from the red-sequence method of SpARCS cluster survey,  \cite{Willis2018temp} find that 10/10 spectroscopically confirmed XMM-LSS clusters at 0.8$\,<\,z\,<\,$1.2 are matched to a SpARCS cluster at the same redshift range. \cite{Foltz2015} observed no measurable differences between the  properties of quiescent cluster members in clusters selected by different methods at $z$\,$\sim$\,1, and they found a remarkable agreement in color-magnitude relation zero point throughout their cluster samples, indicating that the red-sequence method does not preferentially select older, more evolved systems. Furthermore, we note that the $f_{SF}$ in the field in much lower than 1 for high-mass galaxies at all redshifts (Figure \ref{fig:fuvj_vs_z}), so even clusters that show no additional quenching above what would be expected for a field population would still be detected using the red-sequence method.

\subsubsection{Cluster sample: progenitor bin}
\label{sec:Lins}
\cite{Lin2017temp} probed a similar redshift range using a sample of clusters constructed with another red-sequence algorithm, and concluded that the top N clusters from redshift bins occupying similar comoving volumes are a good approach for studying the evolution of clusters and their galaxy populations. Our data allow us to examine the same redshift bins as in \cite{Lin2017temp}'s work ($z$\,=\,0.3-0.6, 0.6-0.77, 0.77-0.9, 0.9-1.02) and select the top N\,=\,30 halos to probe the solidity of our results. We obtain compatible results: the $f_{SF}$ reference value (horizontal lines in Figure \ref{fig:fuvj_vs_rr200}) remained practically unaltered, with variations of around 3\%. The larger differences appear at the highest mass bin of the 0.6 to 0.77 redshift bin, where $f_{SF}$ values  at $R/R_{200}$\,$\sim$\,1.0 are higher and also the field $f_{SF}$ value is around 20\% higher, which could be explained by the slightly narrower and higher redshift bin, since $f_{SF}$ values increase with redshift. In general, the overall shape of the SF fraction as a function of $R/R_{200}$ is almost indistinguishable from that of using the whole sample.

\subsubsection{Cluster center}
As we are analyzing the effect of the environment as measured by the distance to the center of the cluster, normalized by its $R_{200}$ value, the definition of that center could affect our results. Thus, we have also tested the dependence of our results on the identification of the center of the cluster. We define the cluster center as the center of the overdensity of galaxies, as estimated by SpARCS red-sequence algorithm, but we can also calculate the distance to the BCG and use this as a proxy for the $R/R_{200}$ parameter. We can safely conclude that there are no systematics due to the exact definition of the center of the cluster, as using the BCGs does not change our results on the radial dependence of the SF fraction. We observe comparable shapes in all the projections of the $f_{SF}$, $R/R_{200}$, M$_{\star}$, and $z$ parameter space, showing no environmental quenching beyond $
R/R_{200}$\,$\sim$\,1.5 in the redshift range we cover.

\subsubsection{UVJ colors}
The UVJ diagram exhibits a strong dependence on the stellar mass: smaller (bluer) $U-V$ values are populated by lower stellar mass galaxies, while the highest (reddest) $U-V$ values are populated by the most massive galaxies. Thus, our results, especially when binning in stellar mass, could strongly depend on the UVJ boundaries selected (see Section\,\ref{sec:mass_uvj}). We have observed that slight variations in the UVJ limits for classifying SF/quiescent galaxies change the reference value of the SF fraction but not its shape; i.e., the horizontal line in Figure \ref{fig:fuvj_vs_rr200} moves up/down as we shrink/spread the limits with respect to the passive population. The highest stellar mass bin is particularly sensitive to the $V-J$ threshold, and using threshold values larger than 1.6 makes the $f_{SF}$ zero even for larger cluster-centric distances. Nevertheless, we note that the radial gradient of the SF fraction is always apparent, independently of the UVJ limits used.

\section{Discussion}

Considering ``mass quenching'' as part of the secular evolution of a galaxy, we identify the field $f_{SF}$ as the result of the processes related to the ``nature'' of a galaxy. On top of that fiducial $f_{SF}$ value we measure the effect of the cluster environment, which can be interpreted as an acceleration of the quenching process, the ``nurture'' component. In this scenario, we analyze the dependence of the $f_{SF}$ on galaxies' stellar mass and position within the cluster halo, and how it evolves. 

\subsection{The dependence of the $f_{SF}$ on galaxy stellar mass}
Focusing on the $f_{SF}$ curves representing the field environment in Figure \ref{fig:fuvj_vs_m}, we observe how the mass quenching pushes down the $f_{SF}$ starting with the high mass galaxies at earlier epochs and affecting increasingly lower stellar mass galaxies as we move to more recent epochs. 
We introduce the parameters $M^{*}_{\mathrm{INI}}$ and $M^{*}_{\mathrm{END}}$ (Section\,\ref{sec:fsf_m_and_env}) as a direct way to quantify the dependence of $f_{SF}$ on stellar mass.
We are witnessing the mass-dependent evolution of the quenched population: more massive galaxies turn off their star formation becoming red and passive first, as have been observed by many others \citep[e.g.,][]{Cowie1996,Marchesini2009,Muzzin2013,Tomczak2014,Sobral2014}.
Our approach allows us to determine the $f_{SF}$ dependence on stellar mass for different cluster-centric radii and for different redshift bins.  In the same way as described by \cite{Fang2018temp} through their UVJ diagrams grid, we clearly see the buildup of the quiescent population in both the cluster and the field in Figure \ref{fig:fuvj_vs_rr200}. By choosing a stellar mass and redshift bin and searching for the most similar plot at higher mass, we will always identify it with that of an earlier epoch (e.g., panel 9.5\,$\leq$\,log\,$M_{\star}$/M$_{\odot}$\,$<$\,10 at 0.5\,$\leq$\,$z$\,$<$0.7 is most similar to panel 10\,$\leq$\,log\,$M_{\star}$/M$_{\odot}$\,$<$\,10.5 at 0.9\,$\leq$\,$z$\,$<$1.1). 

The rise of $M^{*}_{\mathrm{END}}$ (and $M^{*}_{\mathrm{INI}}$) with redshift also confirms that more massive galaxies evolve more quickly (Figure \ref{fig:mini_mend_vs_z}); while log\,M$_{\star}$/M$_{\odot}$\,$>$\,11.1 galaxies are already mostly quenched by z\,$\sim$\,1, most of the log\,$M_{\star}$/M$_{\odot}$\,$\gtrsim$\,10.6 galaxies are also quiescent by $z$\,$\sim\,$0.4. The $M^{*}_{\mathrm{END}}$ value we have defined is equivalent to the upper stellar mass limit of the blue cloud recently measured by \cite{Haines2017} using the d4000\,-\,$M_{\star}$ plane for field galaxies. The authors find that the upper stellar mass limit of the blue population is steadily retreating towards lower redshifts, from log\,$M_{\star}$/M$_{\odot}$\,=\,11.2 at z\,$\sim$\,0.9 to log\,$M_{\star}$/M$_{\odot}$\,=\,10.9 at z\,$\sim$\,0.58. This is in  good agreement with our field $M^{*}_{\mathrm{END}}$ values. Remarkably, the trend we observe for field $M^{*}_{\mathrm{INI}}$ is also compatible with the local ($z$\,$<$\,0.055) study of \cite{Geha2012}, since these authors find that quenched galaxies in the field do not exist below log\,M$_{\star}$/M$_{\odot}$\,$=$\,9. 

Our quantitative measures of field $M^{*}_{\mathrm{INI}}$ and $M^{*}_{\mathrm{END}}$ for different redshifts could be useful to test galaxy formation models and constrain the physical mechanisms responsible of mass quenching at 0.3\,$\leq$\,$z$\,$<$1.1. These kinds of estimates demonstrate the capability of photometric-redshift studies to produce useful information by providing sample sizes that are much larger than those from spectroscopic surveys over a wide redshift range.

\subsection{The dependence of the $f_{SF}$ on environment}
\subsubsection{Radial gradient of $f_{SF}$}
On top of the ``natural'' buildup of the quiescent population, we observe an excess of quiescent galaxies towards the cluster core that gives evidence to a quenching process driven by the cluster environment. Many previous works have shown the radial dependence of the red/blue fraction in groups and clusters of galaxies. Using a sample of $\sim$\,1000 clusters at 0.45\,$\leq\,z\,\leq$\,0.9 drawn from the first Red-Sequence Cluster Survey (RCS1), \cite{Loh2008} found that the red fraction of $R/R_{200}<$\,0.25 galaxies was much greater than at larger radii and that there is a moderate evolution in the red fraction at radii smaller than 0.5\,$R_{200}$ over the small redshift range. \cite{Raichoor2012}, using an X-ray selected cluster sample of 25 clusters up to z\,$\sim$\,1, showed that the dependence of the fraction of blue galaxies with cluster-centric distances held for different galaxy mass bins. More recently, \cite{Lin2014} and \cite{Jian2017} studied galaxy groups in Pan-STARRS1 Medium-Deep Survey and found that the quiescent fraction slightly decreased as the radius increased, and that the slope of $f_{Q}$ (quiescent fraction) as a function of the cluster-centric radius was steeper for less massive galaxies (log\,$M_{\star}$/M$_{\odot}$\,$\lesssim$\,10.1). Our work confirms and extends these findings to lower stellar mass bins. 

We observe a confinement of the cluster effect to within $R$\,$\simeq$\,$R_{200}$. From panels in Figure \ref{fig:fuvj_vs_rr200}, we can infer that the effect is mostly interior to $R/R_{200}\,\sim\,$0.5, except for the lowest redshift bin or the highest stellar mass bin, where the effect of the environment extends to $R/R_{200}\,\sim\,$1\,-\,1.5. This is also observable in Figure \ref{fig:fuvj_vs_m}, where the cluster $f_{SF}$ vs. log\,$M_{\star}$/M$_{\odot}$ curves show that the strongest impact happens at the very cluster core, while the outskirts curves show the cluster environment affects larger $R/R_{200}$ primarily toward later epochs. \cite{Allen2016} found a similar extent of the cluster effect: cluster SF galaxies within 0.5\,$R_{vir}$ have lower mass-normalized average sizes, and a higher fraction of S\'ersic indices with $n$\,$>$\,1, than field SF galaxies. However, they measure consistent mass-normalized average SFR for SF galaxies in the cluster outskirts and core, but is elevated by a factor of two in the field, suggesting that environmental effects extend to 2\,$R_{vir}$. The authors argue that galaxies in the cluster outskirts have begun to quench, but that is not yet reflected in their sizes/morphologies and they remain star-forming for 2-4\,Gyr after their first cluster infall, in accordance with \cite{Wetzel2013}. The $f_{SF}$ parameter used in this study is sensitive effectively only to completely quenched galaxies. Our result of the radial gradient and our data are equally consistent with two scenarios. The radial gradient of $f_{SF}$, flat from the field to $R$\,$\lesssim$\,$R_{200}$ and then strongly decreasing towards the cluster core, is consistent with a delayed environmental quenching scenario in which galaxies remain unaltered after first entering the cluster outskirts and then rapidly quench when approaching the core. However, the observed radial gradient is also consistent with a scenario in which galaxies after entering the outskirts remain star forming, but at a lower mass-normalized rate, and completed quenching does not occur until significantly well past the virial radius.

From Figure \ref{fig:fuvj_vs_rr200} alone, we cannot definitively deduce whether the extent of the effect of the environment, as indicated by the $R/R_{200}$ at which the $f_{SF}$ drops, evolves or not. The larger extent of the cluster effect at our lowest redshift could be a sign of environmental quenching evolution, but also could be caused by the cluster sample selection. In re-analyzing the data using the \cite{Lin2017temp} cluster sample setup (see \S\,\ref{sec:Lins}),  the differences between the curves in Figure \ref{fig:fuvj_vs_m} representing the field and the cluster outskirts still remain for clusters in the two lower redshift bins ($z\,<$\,0.77). Further analysis of Figure \ref{fig:fuvj_vs_rr200} also gives a hint of a potential dependence of the cluster sphere of influence with galaxies' stellar mass. In the next paper of this series we will investigate such evidence and its implications on the environmental quenching effect.

\subsubsection{The environmental dependence of $M^{*}_{\mathrm{INI}}$ and $M^{*}_{\mathrm{END}}$}
As shown in Figure \ref{fig:mini_mend_vs_z}, the stellar masses $M^{*}_{\mathrm{INI}}$ and $M^{*}_{\mathrm{END}}$ are lower in the cluster core than in the field at all redshifts. This shift indicates that galaxies in the cluster core have their quenching accelerated over what is expected from mass quenching in the field, i.e., cluster galaxies of certain stellar mass complete their quenching earlier than field galaxies of the same stellar mass. Such a shift is at around 0.3-0.4 in $log\,M_{\star}/M_{\odot}$, except for $M^{*}_{\mathrm{INI}}$ at the highest redshift bin where it is reduced to 0.14 and for the $M^{*}_{\mathrm{END}}$ at the lower redshift bins where it is increased to 0.8-0.9. This effect produces $f_{SF}$ vs. $M_{\star}$ curves in \ref{fig:fuvj_vs_m} that are bunched up at the highest redshift bin and more spread out at lower redshifts.
While we observe evolution with redshift for both characteristic masses in all environments, the difference between them, which is a proxy for the slope of the $f_{SF}$ vs. $M_{\star}$ curves, is almost constant for the field and the outskirts of the cluster. For the cluster core, it appears that the $f_{SF}$ vs. $M_{\star}$ curves could be flattening towards lower redshifts, but we are limited by the stellar mass completeness of our two high redshift bins to arrive at a more definitive result.

\subsubsection{The effect of the local galaxy density}
The global environment is sensitive to the cluster-centric radius parameter, which gives the position with respect to the gravitational potential of the parent halo. Thus, $R/R_{200}$ is useful to evaluate the role of mechanisms such as ram-pressure stripping, strangulation, and global tidal effects from the cluster dark matter halo mass. In order to have a complete picture of the quenching processes, one also needs to parametrize the local galaxy density to measure the more immediate environment. Such a parameter will be sensitive to mechanisms that are dependent on the existence of neighboring galaxies, e.g., harassment, galaxy-galaxy interactions, and mergers. 

While our sample size is large enough to study the effect of the stellar mass and the position within the cluster separately, a similar careful analysis of the local galaxy density, at fixed cluster-centric radii, particularly in the cluster core, requires a larger sample of clusters. \cite{Li2012} studied the evolution of the red fraction in a sample of 905 galaxy groups and clusters with 0.15\,$\leq\,z\,<\,$0.52 from the first Red-Sequence Cluster Survey. They analyze the $f_{red}$ dependence on four parameters: galaxy stellar mass, total group stellar mass, normalized group-centric radius, and local galaxy density; and find that the relative effect of local density on $f_{red}$ is roughly constant as a function of group-centric radius, suggesting that one can separate the influence between local galaxy density and the global effects of the group environment. Their results also indicate that both local galaxy density and cluster/group-centric radius are significantly correlated with galaxy population. Therefore, even though a detailed study of the effect of local galaxy density is essential to a clearer picture of galaxy evolution, we expect that the conclusions derived from the analysis of the cluster-centric distances are correct.

\subsection{The evolution of the $f_{SF}$}

The picture that galaxies evolve with redshift has been largely demonstrated since \cite{Butcher1984} showed that the fraction of blue galaxies in clusters is higher at earlier epochs (see \S\,\ref{sec:boeffect}). Our large sample and the use of a large cluster-centric reference $f_{SF}$ (i.e., ``field'') allow us to break down the dependence of the evolution of $f_{SF}$ in detail by measuring $f_{SF}$ vs. $z$ for different stellar mass and cluster-centric radius bins simultaneously.

\cite{Raichoor2012} already showed the benefit of using narrow stellar mass bins when studying the evolution of the blue fraction in clusters. In particular, in their highest stellar mass bin (log\,$M_{\star}$/M$_{\odot}>\,$10.5, the same as in \cite{Butcher1984}'s work), these authors observed no evolution of the blue fraction up to $z\,=\,$0.43, and only when including one cluster at $z$\,$=$\,1.05 they measured an increase of the blue population. Our data allow us to study in detail their redshift gap, and locate the rise of the $f_{SF}$ of the most massive galaxies at $z\,>$\,0.7. Although we cannot make a direct comparison, since their sample of clusters is X-ray selected and their inner cluster bin is up to $R/R_{200}\,=\,$0.5, our results indicate an increment of the $f_{SF}$ with redshift and towards lower stellar masses shallower than the extrapolation of their model (left panels in their Figure 13). 

Figure \ref{fig:fuvj_vs_z} shows that the galaxy evolution observed in clusters, reflected by the increase of the blue population towards higher redshifts, occurs as an extension of the galaxy evolution of the field galaxies. This is not surprising considering that clusters of galaxies are continually accreting galaxies from the field, and that a non-negligible population of field galaxies is blended into the cluster population.
This evolutionary trend is also easily seen from Figure \ref{fig:fuvj_vs_m}. We have shown that the total fraction of the star-forming population is driven by lower-mass galaxies; however, the stellar mass at which galaxies begin to quench is not constant but evolves to higher values with higher redshift, as demonstrated by $M^{*}_{\mathrm{INI}}$. This means that at a fixed stellar mass range, as when calculating $f_{SF}$ of a mass selected sample, we will obtain higher $f_{SF}$ values for higher redshifts. The $f_{SF}$ vs. $z$ relation may appear to be flat if the stellar mass range in which it is computed is selected such that the upper limit is $\lesssim$\,$M^{*}_{\mathrm{INI}}$ of the lowest redshift bin or the lower limit is $\gtrsim$\,$M^{*}_{\mathrm{END}}$ of the highest redshift bin. Consequently, if a study is not careful with the stellar mass limit, one may overestimate the effect (e.g., using a blue luminosity limit instead of a stellar mass limit), or not see any evolutionary effect (e.g., if the stellar mass limit is too high for the redshift range studied).


\begin{figure*}
\centering
\includegraphics[width=\hsize]{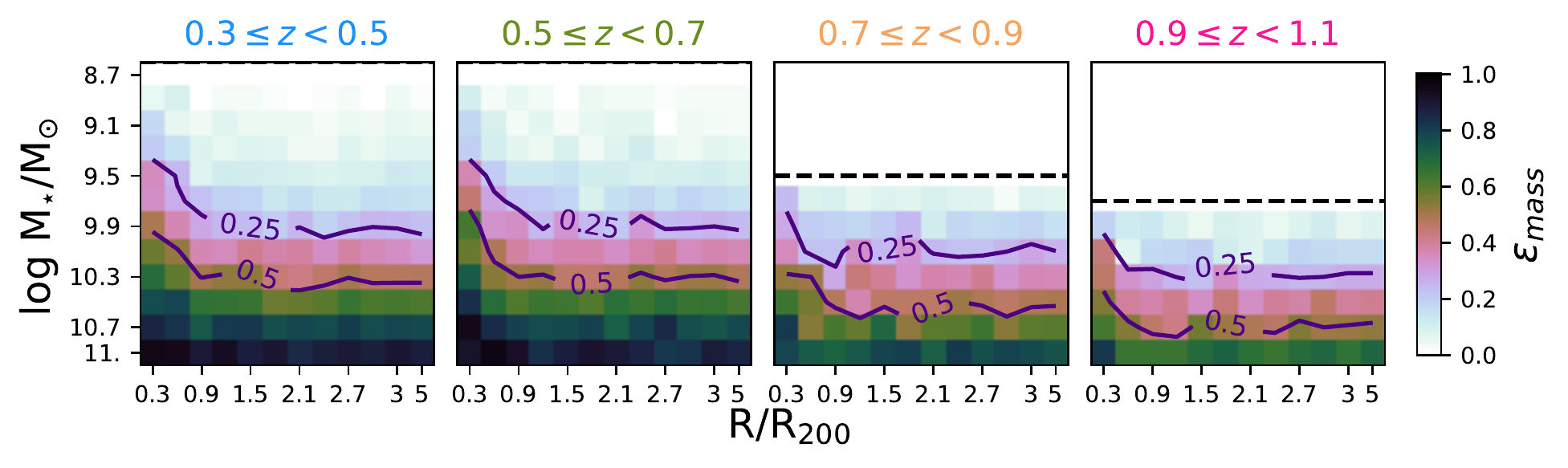}
\caption{Mass quenching efficiency as a function of the stellar mass and the cluster-centric distance for four redshift bins, as indicated by the top labels. Purple curves show the contours at fixed $\epsilon_{mass}$\,=\,0.25 and 0.5. Horizontal dashed lines indicate the stellar mass completeness limit of each redshift bin.}
\label{fig:mass_quench_2d}
\end{figure*}

\begin{figure*}
\centering
\includegraphics[width=\hsize]{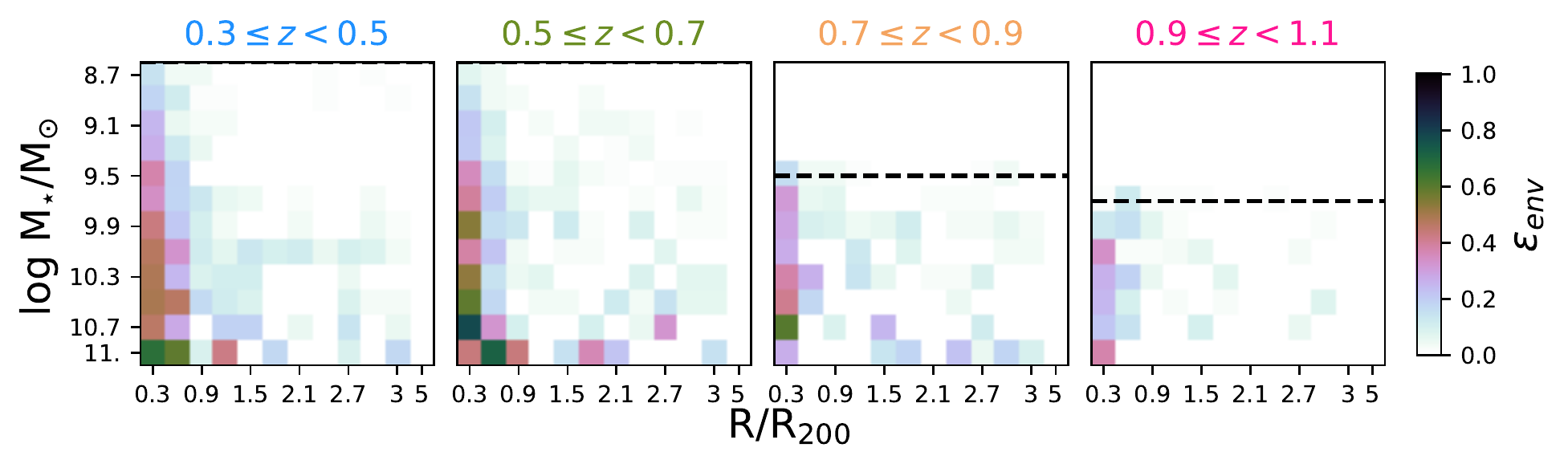}
\caption{Environmental quenching efficiency as a function of the stellar mass and the cluster-centric distance for four redshift bins, as indicated by the top labels. Horizontal dashed lines indicate the stellar mass completeness limit of each redshift bin.}
\label{fig:env_quench_2d}
\end{figure*}

\begin{figure*}
\centering
\includegraphics[width=\hsize]{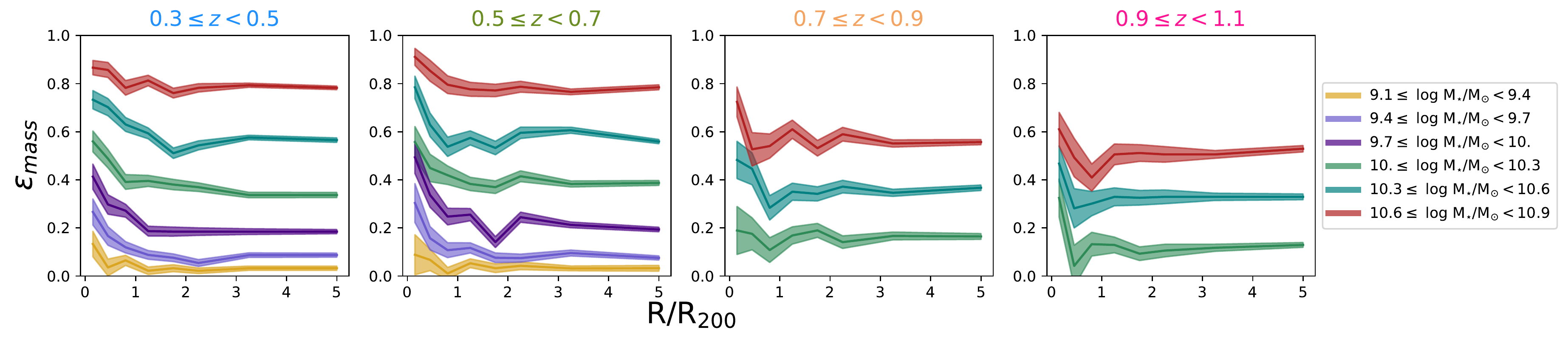}
\caption{Mass quenching efficiencies as a function of cluster-centric distance for four redshift bins, as indicated by the top labels, and different stellar mass bins.}
\label{fig:mass_quench_vs_env}
\end{figure*}


\subsection{Environmental and mass quenching efficiencies}

\subsubsection{Measuring $\epsilon_{env}$ and $\epsilon_{mass}$}
Following the approach of previous works, we estimate the environmental quenching efficiency, $\epsilon_{env}$, to explore the degree to which cluster galaxies have suppressed their star formation relative to similar galaxies in the field \citep{vandenBosch2008,Peng2010,Phillips2014}. The formula that we employ is $\epsilon_{env}(r,m)\,=\,(f_{Q}(r,m)\,-\,f_{Q}(r_{field},m))/f_{SF}(r_{field},m)$, which accounts for the fraction of galaxies with stellar mass $m$ at a certain distance $r$ from the cluster in excess of those in the field. As in previous sections, the field environment is defined as between $R/R_{200}=\,$4 and 6. In the same way, the stellar mass quenching efficiency, $\epsilon_{mass}$, is defined as the fraction of galaxies that are quenched compared to the SF population at low stellar mass: $\epsilon_{mass}(r,m)\,=\,(f_{Q}(r,m)\,-\,f_{Q}(r,m_0))/f_{SF}(r,m_0)$, where the reference mass $m_{0}$ is the lowest stellar mass at which (almost) all galaxies at cluster-centric radius $r$ are forming stars. In practice, we use the stellar mass completeness limit for each redshift bin. 
While these definitions are usually employed when studying clusters, $\epsilon_{env}$ and $\epsilon_{mass}$ are usually defined by means of the local galaxy density when analyzing field galaxies. For $\epsilon_{env}$, works in the literature usually compare the top local galaxy density quartile to the bottom quartile; e.g., \cite{Kawinwanichakij2017} use $\delta_{75}$ and $\delta_{25}$ instead of $r_{cluster}$ and $r_{field}$. Similarly, $\epsilon_{mass}$ is defined in field studies avoiding galaxies within the densest environments, e.g., using only galaxies in the three bottom local galaxy density quartiles, $\delta<\delta_{75}$. These different definitions complicate direct comparisons between different works and the resulting interpretations.
 
In Figures \ref{fig:mass_quench_2d} and \ref{fig:env_quench_2d}  we plot the 2D profile of the two quenching efficiencies as a function of galaxy stellar mass and cluster-centric radius for different redshift bins. We note that $\epsilon_{env}$ and $\epsilon_{mass}$ are degenerate as the fraction of SF galaxies, $f_{SF}(r,m)$, approaches 0, as both parameters would approach the value of 1. When (almost) all galaxies are quenched we cannot distinguish which of the two quenching mechanisms, mass or environment, has been more efficient.

\subsubsection{Are mass and environmental quenching efficiencies independent?}
\label{sec:mass_env_quench_indep}

In the case of $\epsilon_{mass}$ being independent of the environment we expect in Figure \ref{fig:mass_quench_2d} horizontal stripes with a higher mass quenching efficiency for more massive galaxies and lower efficiency towards lower stellar masses. We do observe a gradient with stellar mass, but the stripes are not completely flat. This effect is more clear in Figure \ref{fig:mass_quench_vs_env}, where the mass quenching efficiency is plotted as a function of $R/R_{200}$ for various stellar mass bins. We can see how $\epsilon_{mass}$ appears as flat lines on top of each other as we move towards higher stellar mass bins for environments at $R\,\gtrsim\,R_{200}$, indicating that $\epsilon_{mass}$ is independent of $R/R_{200}$. \cite{Peng2010}, using galaxy samples from SDSS and zCOSMOS concluded that $\epsilon_{mass}$ is completely independent of the environment, where environment is defined as quartiles in local density.
However, as we approach the cluster core, a dependence of $\epsilon_{mass}$ on $R/R_{200}$ becomes evident: $\epsilon_{mass}$ is clearly more efficient towards the cluster center. This fact suggests that in very dense areas, such as the cores of clusters, galaxy formation history and evolution may be different from that of the field. We may expect galaxies in the cluster core, on average, to be older than those in the field, since some of them are likely formed ``in situ'' \citep[e.g.,][]{Maulbetsch2007}. These galaxies may be formed earlier and it is reasonable that they follow a different mass quenching history than that of the ``universal clock'' of field galaxy mass quenching. \cite{Peng2010} likely did not find this population of galaxies formed in very dense environments because their galaxy density did not sample sufficiently dense environments, or the effect was masked by the large uncertainty in estimating local galaxy density.

Looking at Figures \ref{fig:mass_quench_2d} and \ref{fig:env_quench_2d}, we see how $\epsilon_{mass}$ clearly dominates over $\epsilon_{env}$ outside $R$\,$\sim$\,$R_{200}$. But inside $R_{200}$, $\epsilon_{env}$ grows faster than $\epsilon_{mass}$, such that for the intermediate to low stellar mass range ($M_{\star}\lesssim\,$10$^{10.3}$-10$^{9.7}$M$_{\odot}$ depending on the redshift) very close to the cluster center they appear to be equally effective. 
We also observe how both effects become less efficient with decreasing stellar masses, in agreement with, e.g., \cite{Jian2017}. Such a clear dependence of $\epsilon_{env}$ on stellar mass has been reported by \cite{Kawinwanichakij2017} at $z$\,$>$\,1. These authors found that more massive galaxies experience stronger environmental quenching, and they concluded that the environmental quenching effects are not separable from stellar mass at $z$\,$\gtrsim$\,1. However, at 0.5\,$<$\,$z$\,$<$\,1, they measured a $\epsilon_{env}$ nearly independent of stellar mass. \cite{vdBurg2018temp} found also that there is no clear stellar-mass dependence of $\epsilon_{env}$ in any radial bin for a sample of \textit{Planck}-selected clusters at 0.5\,$<$\,$z$\,$<$\,0.7. However, the stellar mass dependence that we find in this redshift bin is mainly observed below their lowest stellar mass bin, i.e., $\lesssim$\,10$^{9.7}$M$_{\odot}$. We also note that their sample of clusters consist of the most massive halos, very likely more massive than the average of the halos we are probing. \cite{Papovich2018}, using the same ZFOURGE dataset as \cite{Kawinwanichakij2017}, studied the stellar mass function of the quiescent population and deduced that the observed shape requires an environmental quenching that depends on stellar mass even for the lowest redshift bin of 0.5\,$<$\,$z$\,$<$\,1. Our measurement of the environmental quenching is consistent with their deduction, and extends to lower redshifts and the very dense environment of the cluster cores.

Our results suggest that, within the cores of clusters, mass and environmental quenching efficiencies, as defined, are inter-dependent, and not separable. However, the environmental influence in the core of clusters is unlikely to arise from simply the much higher local galaxy density, as the gravitational potential of the parent halo must also exert a significant influence. In future work, we will study this in greater detail by isolating the effects of local galaxy density at different cluster centric radii, as attempted by \citep{Li2012}, which may provide key insights into the possible different quenching mechanisms at play.

\begin{figure}
\centering
\includegraphics[width=\hsize]{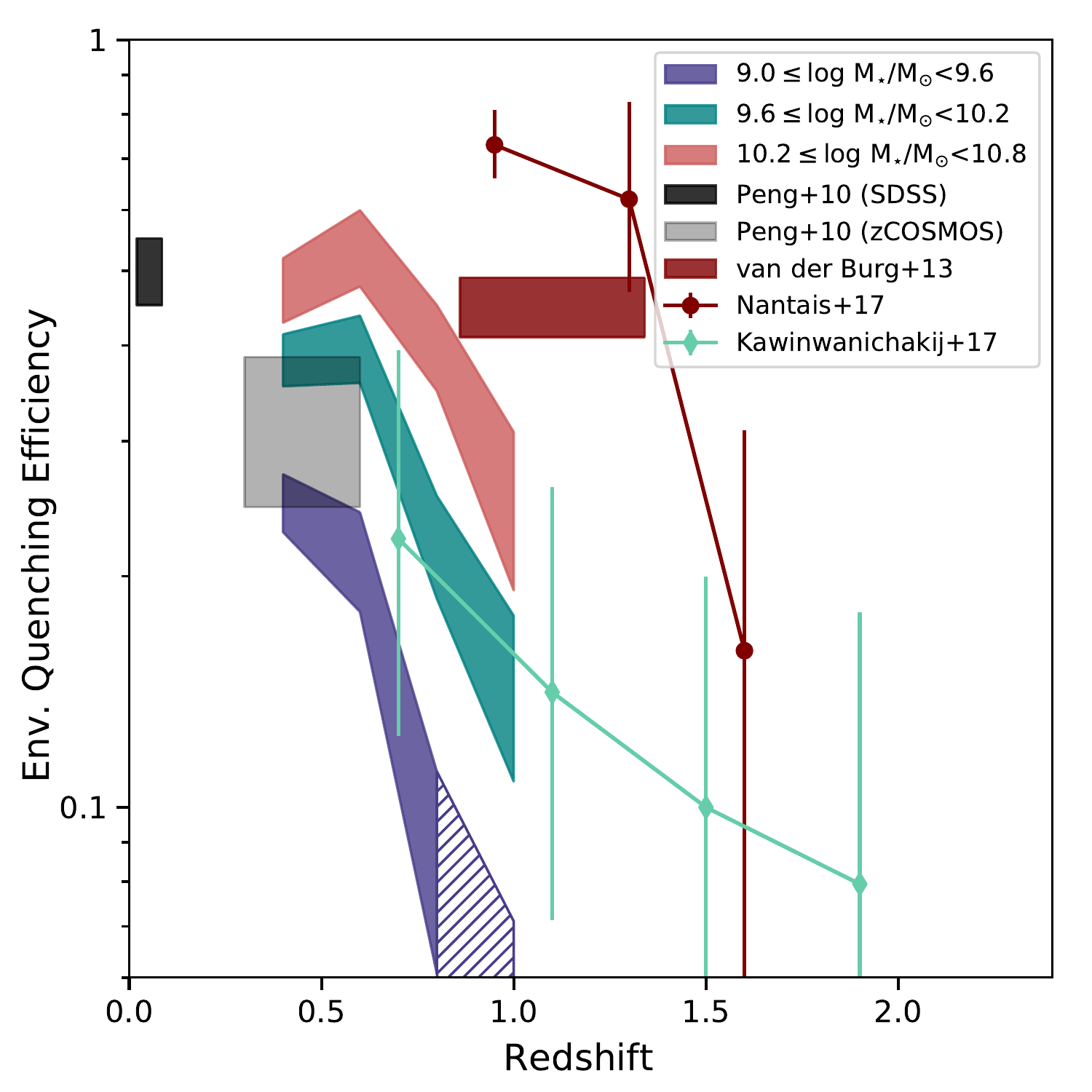}
\caption{Environmental quenching efficiency as a function of redshift for different stellar mass bins (red, teal, and indigo regions) in the cores ($R/R_{200}<\,$0.4) of our cluster sample and various other samples from the literature. Hatched area indicates that the stellar mass is below our stellar mass completeness limit.
Black and gray regions show the environmental quenching efficiency of galaxies from \cite{Peng2010} for their highest galaxy density quartile for galaxies with 9.0\,$<$\,log\,M$_{\star}$/M$_{\odot}<$\,11.0 from SDSS and with 10.2\,$<$\,log\,M$_{\star}$/M$_{\odot}<$\,11.0 from zCOSMOS, respectively. Red circles show the measurements from \cite{Nantais2017} for galaxies in clusters with log\,M$_{\star}$/M$_{\odot}<$\,10.3, while the red rectangle region shows the result of \cite{vanderBurg2013} using the same sample up to $z\,=\,$1.34. Turquoise circles show the intermediate stellar mass bin (9.8\,$<$\,log\,M$_{\star}$/M$_{\odot}<$\,10.2) of galaxies in the highest-density quartile from \cite{Kawinwanichakij2017}.}
\label{fig:env_quench_z}
\end{figure}

\begin{figure}
\centering
\includegraphics[width=\hsize]{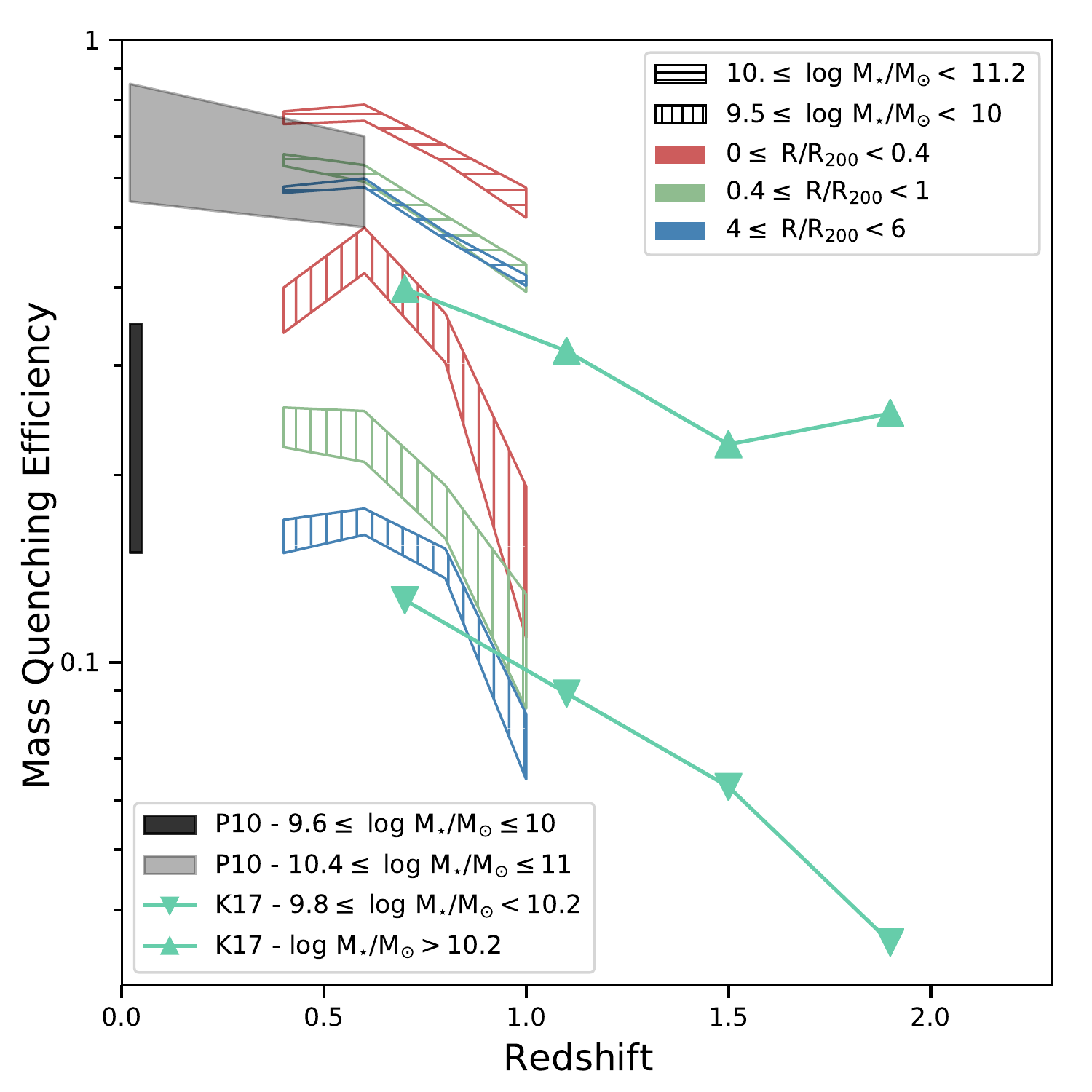}
\caption{Mass quenching efficiency as a function of redshift for three $R/R_{200}$ bins (red, green, and blue regions) and two stellar mass bins (vertical and horizontal hatched areas) for our sample of clusters. Gray and black regions show the field mass quenching efficiency of galaxies in two stellar mass bins from \cite{Peng2010}'s results in SDSS and zCOSMOS. Turquoise up and down triangles show the high and intermediate stellar mass bins, respectively, from \cite{Kawinwanichakij2017}, whose mass quenching efficiencies are computed for galaxies in local galaxy densities below the 75th percentile ($\delta\,<\,\delta_{75}$).}
\label{fig:mass_quench_z}
\end{figure}

\subsubsection{The evolution of the environmental quenching efficiency}
From the $\epsilon_{env}$ displayed in Figure \ref{fig:env_quench_2d}, we observe how the influence of the cluster diminishes with increasing redshift. At z\,$>$\,0.7, $\epsilon_{env}$ becomes significant only at $R/R_{200}\lesssim$\,0.6. In Figure \ref{fig:env_quench_z} we show the evolution of the environmental quenching efficiency in the cluster core ($R/R_{200}<$\,0.4) and its stellar mass dependence. By estimating $\epsilon_{env}$ in three stellar mass bins, we see  how the environment is more effective at quenching more massive galaxies over the whole redshift range, as clearly revealed by the color $\epsilon_{env}$ map of Figure \ref{fig:env_quench_2d}.
Figure \ref{fig:env_quench_z} shows a strong evolution of the environmental quenching in our redshift range of study, for all stellar masses. We also observe that such evolution of $\epsilon_{env}$ is similar for different stellar masses. 
By definition, $\epsilon_{env}$ accounts for the fraction of galaxies that have been quenched by the cluster in excess of those in the field across all time. Hence the evolution with redshift we measure for $\epsilon_{env}$ could just reflect the increased time that galaxies have been in clusters by $z$\,$\sim$\,0.4 compared to $z$\,$\sim$\,1. However, to attribute the observed evolution to this, the fading time scale needs to be longer than the infall time scale.  
Therefore, the changes we see in $\epsilon_{env}$ with redshift likely reflect a true evolution of the environmental quenching efficiency.

Due to systematic effects such as stellar mass completeness limits, cluster sample selection, and different definitions of environment (e.g., local galaxy density, different $R/R_{200}$ distances), it is difficult to compare our $\epsilon_{env}$ curves with those in the literature directly.
For field galaxies, \cite{Peng2010} and \cite{Kawinwanichakij2017} show lower $\epsilon_{env}$ values and a smoother evolution as illustrated in Figure \ref{fig:env_quench_z}. These studies define environment through local galaxy density and they do not focus on galaxy clusters; so, our $f_{SF}(r_{cluster},m)$ estimate is very likely measured in a much denser environment than the highest density quartile that \cite{Kawinwanichakij2017} used to calculate $\epsilon_{env}$. Conversely, our $f_{SF}(r_{field},m)$ reference value is measured as the average between $R/R_{200}=$\,4 and 6, which is likely closer to the 50th percentile in local density than the lowest density quartile used by \cite{Kawinwanichakij2017}. This difference would at least partly compensate for the $f_{SF}(r_{cluster},m)$ measured in a denser environment; therefore, if we could use a $f_{SF}(r_{field},m)$ comparable to that of \cite{Kawinwanichakij2017}, we would expect our data to show an even higher $\epsilon_{env}$. Note that we only include \cite{Kawinwanichakij2017}'s intermediate mass bin in Figure \,\ref{fig:env_quench_z}. The steeper slopes of our curves compared to \cite{Peng2010}'s and \cite{Kawinwanichakij2017}'s trends could be interpreted as the acceleration of the quenching produced by the cluster with respect to the field/group environment.

\cite{Nantais2017} analyzed a sample of spectroscopically confirmed SpARCS galaxy clusters at 0.86\,$<$\,z\,$<$\,1.65 with log\,M$_{\star}$/M$_{\odot}$\,$>$\,10.3, showing a substantial evolution of the environmental quenching efficiency within their redshift range. Their trend is compatible with the strong evolution that we observe (Figure \ref{fig:env_quench_z}), though the $\epsilon_{env}$ we measure for our highest stellar mass bin in the overlapping redshift bin at z\,$\sim$\,1 is considerably lower. However, our measurements are in better agreement with \cite{vanderBurg2013} using the same data as \cite{Nantais2017} at 0.86\,$<$\,$z$\,$<$\,1.34. The large discrepancy is very likely due to how the environmental quenching efficiency is computed, e.g., \cite{Nantais2017} used different stellar mass completeness limits than \cite{vanderBurg2013}. Furthermore, we define the cluster core within $R/R_{200}$\,$\leq$\,0.4 and use the same dataset to account for the field, while \cite{Nantais2017} and \cite{vanderBurg2013} consider all galaxies within 1\,Mpc and their field galaxies are from the UltraVISTA/COSMOS catalog.

\subsubsection{The evolution of the mass quenching efficiency}

In Figure \ref{fig:mass_quench_z} we focus on the evolution of the mass quenching efficiency in three different environments: the cluster core ($R/R_{200}$\,$<$\,0.4), the outskirts of the cluster (0.4\,$\leq$\,$R/R_{200}$\,$<$\,1), and the field (4\,$\leq$\,$R/R_{200}$\,$<$\,6). The mass dependence is also analyzed by using two stellar mass bins. As with the environmental quenching, the efficiency of the mass quenching diminishes with higher redshift. We observe how for the two stellar mass bins the strongest evolution in our redshift range happens between $z$\,$\sim$\,0.6 and 1. However, the slope of the lower stellar mass bin is steeper, suggesting that the evolution of $\epsilon_{mass}$ is faster for lower stellar masses. The $R/R_{200}$ bins show again that the mass quenching within the cluster environment ($R$\,$\lesssim$\,$R_{200}$) is different from that of the field. While it still depends on the mass of the galaxy, i.e., mass quenching is more efficient for more massive galaxies, galaxies in the cluster core are, on average, in a more advanced stage of mass quenching. The mass quenching dependence on environment within the cluster core and stellar mass is clearly present in our entire redshift range. However, for each mass bin, the slopes of the three $R/R_{200}$ bins are very similar, suggesting that the environment does not affect the evolution of the mass quenching efficiency, consistent with the scenario that there is a population of galaxies formed ``in situ'' in the cluster cores in earlier time, as described in \S\,\ref{sec:mass_env_quench_indep}.

Our results for the outskirts and field environments (blue and green hatched curves in Figure \ref{fig:mass_quench_z}) are in reasonable agreement with previous works on field galaxies. Our two stellar mass field bins are compatible at lower redshift with results from \cite{Peng2010} and at higher redshift with \cite{Kawinwanichakij2017}'s high and intermediate mass bins, both in amplitude and evolutionary rate with redshift. The differences can likely be attributed to the slightly different stellar mass limits, or the difference in stellar mass computation.

The interpretation of redshift evolution is complex due to selection effects and, while the threshold of $N_{\mathrm{red}}$\,$>$\,6.0 takes care of the sensitivity to lower-mass clusters with decreasing redshift, by using a fixed mass limit we are not analyzing a progenitor/descendant sample. However, when using \cite{Lin2017temp}'s setup described in \S\,\ref{sec:Lins}, we obtain very similar evolutionary trends as in Figure \ref{fig:mass_quench_z} (mass-limited sample)
This suggests that the effect of a mass-limited cluster sample does not mask the trend with redshift.

\subsubsection{Clues to the quenching process}

Our results show a sharp decline of $f_{SF}$ at $R$\,$\lesssim$\,$R_{200}$ after a remarkably steady field value, which extends up to 6\,$R/R_{200}$. The constant $f_{SF}$ value from the outskirts to as far as 6\,$R/R_{200}$ indicates that little or no environmental quenching is acting outside the cluster, or if there is an environmental quenching process acting out to 6\,$R/R_{200}$, it is independent of the distance to the cluster center. The latter could be the case if, e.g., pre-processing in groups or high-density regions is occurring. At lower redshifts, several works have found that the fraction of star forming galaxies in clusters is suppressed with respect to the field value even at 3\,$R/R_{200}$ \citep[e.g.,][]{Haines2015,Chung2011}. The comparison with these works is complex since, for example, the sample selection is different, as we select SF galaxies by the UVJ diagram while they use an SFR$_\mathrm{IR}$ threshold. Nevertheless, in the bottom panels of Figure \ref{fig:fuvj_vs_rr200}, we observe a trend of increasing $R/R_{200}$ in the effect of the environment from high to low redshift, which could be compatible with the environmental effect found at 3\,$R/R_{200}$ for z\,$<$\,0.3 clusters.
To come to a more definitive conclusion we need a larger sample that allows us the study of the fraction of SF galaxies as a function of the local galaxy density at fixed cluster-centric radii. If $f_{SF}$ vs. local galaxy density show a correlation that is independent of $R/R_{200}$, at least between $R_{200}$ and 6\,$R_{200}$, pre-processing would be a suitable mechanism for explaining the results.

More interestingly, within the cluster core, we show: (1) that the  $f_{SF}$ vs. $M_{\star}$ curve of the cluster is shifted with respect to the field curve; (2) that $\epsilon_{mass}$ increases at smaller $R$/$R_{200}$; and, (3) that $\epsilon_{env}$ depends on the galaxy stellar mass and evolves with redshift to higher values at later epochs. From (1), we deduce that the cluster environment accelerates the quenching, and (2) indicates that part of the advanced quenching level in the cluster core may be the outcome of an earlier population formed ``in situ'' in the primordial overdensities where clusters are located. Additionally, (3) may indicate that the efficiency of the stellar-mass-dependent physical process(es) responsible for environmental quenching may change with time, being more efficient at lower redshifts. Strangulation/starvation are environmental mechanisms which can produce such a mass-dependent efficiency as galaxies become quenched after consuming all their cold gas reservoir, once their hot gas supply was cut off when falling into the cluster host halo. For example, in \cite{Davies2016}'s low redshift model, starvation is the favorite model for quenching intermediate-mass galaxies. At higher redshift ($z$\,$\gtrsim$\,1), our results are consistent with \cite{Balogh2016}'s model, where the truncation of new gas supply limits the star-forming lifetime of a galaxy as it exhausts its reservoir through star formation and outflows \citep[the ``overconsumption'' model,][]{McGee2014}. In such a model, the higher star formation rates of more massive galaxies lead to shorter quenching times, leading to a $\epsilon_{env}$ that rises with stellar mass. Our results fit in this model and may suggest that, at intermediate redshifts (0.3\,$<$\,$z$\,$<$\,1.1), starvation/strangulation become more efficient at quenching intermediate to more massive galaxies.

\section{Summary and conclusions}

We have analyzed the fraction of star-forming galaxies, $f_{SF}$, in a sample of 209 SpARCS clusters at 0.3\,$\leq$\,z$_{CL}$\,$\leq$\,1.1. The large area and depth of the HSC-SSP data, along with our wide redshift range, has allowed us to study the dependence of the $f_{SF}$ on galaxy stellar mass and distance to the cluster center, and the evolution of these parameters with redshift. Our major results can be summarized as follows:
\begin{enumerate}
\item On a grid of stellar masses and redshifts, we demonstrate that the $f_{SF}$ decreases from $R/R_{200}$\,$\sim$\,1 towards the cluster core beyond that measured in field galaxies of similar stellar masses and redshifts. More massive cluster galaxies show a decrease at earlier redshifts, while lower mass cluster galaxies begin such a decrease at lower redshifts.

\item Using the same dataset to measure the evolution of $f_{SF}$ with redshift in the field and the cluster, we find that the environmental effect produces not just only a shift to lower values but also a faster evolution rate. The relative decrease in $f_{SF}$ between the cluster core and the field  is strongly dependent on the galaxy stellar mass and cluster redshift.

\item We estimate $M^{*}_{\mathrm{INI}}$ and $M^{*}_{\mathrm{END}}$, the fiducial stellar masses at which star-forming galaxies begin to quench and at which almost all galaxies (80\%) are already quenched, respectively, in bins of cluster-centric radii, as a function of redshift. These fiducial masses show a clear evolution in redshift, with lower values towards lower redshifts, indicating that more massive galaxies begin and complete their quenching process earlier. Furthermore, we observe a definitive dependence of $M^{*}_{\mathrm{INI}}$ and $M^{*}_{\mathrm{END}}$ on the environment, as defined by the distance to the cluster core, with both $M^{*}_{\mathrm{INI}}$ and $M^{*}_{\mathrm{END}}$ being lower in the cluster core compared to the field. This strong environmental dependence demonstrates that galaxies of a given mass in clusters begin and complete their quenching of star formation much earlier than their counterparts in the field.

\item We calculate environmental and mass quenching efficiencies as a function of stellar mass and $R/R_{200}$. We find that outside the cluster core, mass and environmental quenching efficiencies are separable. However, in the cluster core $\epsilon_{mass}$ increases by a significant amount relative to that of the field, suggesting that galaxies in the cluster core are in a more advanced stage of mass quenching. This difference may be the result of an initial population of galaxies formed ``in situ'' in the cores of the clusters, well ahead of the general population of field galaxies. Furthermore, the environmental quenching in the cluster core is less efficient for lower mass galaxies. We conclude that, within the cluster core, mass and environmental quenching efficiencies are not separable. We also observe that low-mass galaxies in general have lower $\epsilon_{env}$, suggesting the effect of environmental quenching has a dependence on galaxy mass even outside the cluster core.

\item We find that both mass and environmental quenching processes evolve with redshift, becoming more efficient at later epochs. The evolution of the environmental quenching efficiency is similar for different stellar masses, but more rapid in the cluster core than for the field/group environments. On the other hand, while the mass quenching efficiency also evolves in a similar manner for galaxies in different environments over our redshift range, the evolution of $\epsilon_{mass}$ is more rapid for low mass galaxies.
\end{enumerate}

In this work, we have used cluster-centric radius as an environmental parameter, and have not isolated the effects of local galaxy density simultaneously. Such an investigation can provide key insights into the quenching process and mechanisms. A dataset with a larger sample of clusters is required to carry-out such an analysis. Such a dataset will also allow us to separate the sample into different cluster richness, allowing the examination of the dependence of these results on the cluster mass. A larger redshift baseline with deeper data will provide crucial information on the evolution of the more massive cluster galaxies at higher redshifts and the beginning of the quenching of the less massive galaxies in different epochs. 

\acknowledgments
In memory of Prof. Juan-Luis Pintos: sociologist and beloved father. We thank the anonymous referee for the careful reading of the manuscript and
providing many useful comments. We also thank Remco van der Burg for valuable discussion and constructive feedback on the manuscript.

This work is based on data collected at the Subaru Telescope and retrieved from the HSC data archive system which is operated by Subaru Telescope and Astronomy Data Center at National Astronomical Observatory of Japan. 
This work is also based on observations made with the Spitzer Space Telescope, which is operated by Jet Propulsion Laboratory of the California Institute of Technology under NASA Contract 1407.
This work also uses data obtained at UKIRT, which is owned by the University of Hawaii (UH) and operated by the UH Institute for Astronomy; operations are enabled through the cooperation of the East Asian Observatory.
This work is also based in part on observations obtained with MegaPrime/MegaCam, a joint project of CFHT and CEA/DAPNIA, at the Canada-France-Hawaii Telescope (CFHT) which is operated by the National Research Council (NRC) of Canada, the Institut National des Science de l’Univers of the Centre National de la Recherche Scientifique (CNRS) of France, and the University of Hawaii.

The research presented in this paper is supported by grants from the Discovery Grant Program and the Canada Research Chair Program of the National Science and Engineering Council of Canada (NSERC), and the Faculty of Arts and Science of University of Toronto to H. Yee. L. Old acknowledges a European Space Agency (ESA) Research Fellowship. The research of G. Wilson is supported by the National Science Foundation through grant AST-1517863, by HST program numbers GO-13677/14327.01 and GO-15294, and by grant number 80NSSC17K0019 issued through the NASA Astrophysics Data Analysis Program (ADAP). Support for program numbers GO-13677/14327.01 and GO-15294 was provided by NASA through a grant from the Space Telescope Science Institute, which is operated by the Association of Universities for Research in Astronomy, Incorporated, under NASA contract NAS5-26555. 

This research made use of NumPy, SciPy, Astropy, and Matplotlib.

\bibliographystyle{aasjournal}
\bibliography{references}

\begin{thebibliography}{}
\expandafter\ifx\csname natexlab\endcsname\relax\def\natexlab#1{#1}\fi
\providecommand{\url}[1]{\href{#1}{#1}}

\bibitem[{{Abazajian} {et~al.}(2004){Abazajian}, {Adelman-McCarthy},
  {Ag{\"u}eros}, {Allam}, {Anderson}, {Anderson}, {Annis}, {Bahcall}, {Baldry},
  {Bastian}, {Berlind}, {Bernardi}, {Blanton}, {Bochanski}, {Boroski},
  {Briggs}, {Brinkmann}, {Brunner}, {Budav{\'a}ri}, {Carey}, {Carliles},
  {Castander}, {Connolly}, {Csabai}, {Doi}, {Dong}, {Eisenstein}, {Evans},
  {Fan}, {Finkbeiner}, {Friedman}, {Frieman}, {Fukugita}, {Gal}, {Gillespie},
  {Glazebrook}, {Gray}, {Grebel}, {Gunn}, {Gurbani}, {Hall}, {Hamabe},
  {Harris}, {Harris}, {Harvanek}, {Heckman}, {Hendry}, {Hennessy}, {Hindsley},
  {Hogan}, {Hogg}, {Holmgren}, {Ichikawa}, {Ichikawa}, {Ivezi{\'c}}, {Jester},
  {Johnston}, {Jorgensen}, {Kent}, {Kleinman}, {Knapp}, {Kniazev}, {Kron},
  {Krzesinski}, {Kunszt}, {Kuropatkin}, {Lamb}, {Lampeitl}, {Lee}, {Leger},
  {Li}, {Lin}, {Loh}, {Long}, {Loveday}, {Lupton}, {Malik}, {Margon},
  {Matsubara}, {McGehee}, {McKay}, {Meiksin}, {Munn}, {Nakajima}, {Nash},
  {Neilsen}, {Newberg}, {Newman}, {Nichol}, {Nicinski}, {Nieto-Santisteban},
  {Nitta}, {Okamura}, {O'Mullane}, {Ostriker}, {Owen}, {Padmanabhan},
  {Peoples}, {Pier}, {Pope}, {Quinn}, {Richards}, {Richmond}, {Rix}, {Rockosi},
  {Schlegel}, {Schneider}, {Scranton}, {Sekiguchi}, {Seljak}, {Sergey},
  {Sesar}, {Sheldon}, {Shimasaku}, {Siegmund}, {Silvestri}, {Smith}, {Smol{\v
  c}i{\'c}}, {Snedden}, {Stebbins}, {Stoughton}, {Strauss}, {SubbaRao},
  {Szalay}, {Szapudi}, {Szkody}, {Szokoly}, {Tegmark}, {Teodoro}, {Thakar},
  {Tremonti}, {Tucker}, {Uomoto}, {Vanden Berk}, {Vandenberg}, {Vogeley},
  {Voges}, {Vogt}, {Walkowicz}, {Wang}, {Weinberg}, {West}, {White}, {Wilhite},
  {Xu}, {Yanny}, {Yasuda}, {Yip}, {Yocum}, {York}, {Zehavi}, {Zibetti}, \&
  {Zucker}}]{Abazajian2004}
{Abazajian}, K., {Adelman-McCarthy}, J.~K., {Ag{\"u}eros}, M.~A., {et~al.}
  2004, \aj, 128, 502

\bibitem[{{Adami} {et~al.}(2011){Adami}, {Mazure}, {Pierre}, {Sprimont},
  {Libbrecht}, {Pacaud}, {Clerc}, {Sadibekova}, {Surdej}, {Altieri}, {Duc},
  {Galaz}, {Gueguen}, {Guennou}, {Hertling}, {Ilbert}, {Le F{\`e}vre},
  {Quintana}, {Valtchanov}, {Willis}, {Akiyama}, {Aussel}, {Chiappetti},
  {Detal}, {Garilli}, {Lebrun}, {Lef{\`e}vre}, {Maccagni}, {Melin}, {Ponman},
  {Ricci}, \& {Tresse}}]{Adami2011}
{Adami}, C., {Mazure}, A., {Pierre}, M., {et~al.} 2011, \aap, 526, A18

\bibitem[{{Aihara} {et~al.}(2017){Aihara}, {Armstrong}, {Bickerton}, {Bosch},
  {Coupon}, {Furusawa}, {Hayashi}, {Ikeda}, {Kamata}, {Karoji}, {Kawanomoto},
  {Koike}, {Komiyama}, {Lupton}, {Mineo}, {Miyatake}, {Miyazaki}, {Morokuma},
  {Obuchi}, {Oishi}, {Okura}, {Price}, {Takata}, {Tanaka}, {Tanaka}, {Tanaka},
  {Uchida}, {Uraguchi}, {Utsumi}, {Wang}, {Yamada}, {Yamanoi}, {Yasuda},
  {Arimoto}, {Chiba}, {Finet}, {Fujimori}, {Fujimoto}, {Furusawa}, {Goto},
  {Goulding}, {Gunn}, {Harikane}, {Hattori}, {Hayashi}, {Helminiak}, {Higuchi},
  {Hikage}, {Ho}, {Hsieh}, {Huang}, {Huang}, {Imanishi}, {Iwata}, {Jaelani},
  {Jian}, {Kashikawa}, {Katayama}, {Kojima}, {Konno}, {Koshida}, {Kusakabe},
  {Leauthaud}, {Lee}, {Lin}, {Lin}, {Mandelbaum}, {Matsuoka}, {Medezinski},
  {Miyama}, {Momose}, {More}, {More}, {Mukae}, {Murata}, {Murayama}, {Nagao},
  {Nakata}, {Niikura}, {Nishizawa}, {Oguri}, {Okabe}, {Ono}, {Onodera},
  {Onoue}, {Ouchi}, {Pyo}, {Shibuya}, {Shimasaku}, {Simet}, {Speagle},
  {Spergel}, {Strauss}, {Sugahara}, {Sugiyama}, {Suto}, {Suzuki}, {Tait},
  {Takada}, {Terai}, {Toba}, {Turner}, {Uchiyama}, {Umetsu}, {Urata}, {Usuda},
  {Yeh}, \& {Yuma}}]{Aihara2017}
{Aihara}, H., {Armstrong}, R., {Bickerton}, S., {et~al.} 2017, ArXiv e-prints,
  arXiv:1702.08449

\bibitem[{{Allen} {et~al.}(2016){Allen}, {Kacprzak}, {Glazebrook}, {Tran},
  {Spitler}, {Straatman}, {Cowley}, \& {Nanayakkara}}]{Allen2016}
{Allen}, R.~J., {Kacprzak}, G.~G., {Glazebrook}, K., {et~al.} 2016, \apj, 826,
  60

\bibitem[{{Andreon} {et~al.}(2006){Andreon}, {Quintana}, {Tajer}, {Galaz}, \&
  {Surdej}}]{Andreon2006}
{Andreon}, S., {Quintana}, H., {Tajer}, M., {Galaz}, G., \& {Surdej}, J. 2006,
  \mnras, 365, 915

\bibitem[{{Arnouts} {et~al.}(1999){Arnouts}, {Cristiani}, {Moscardini},
  {Matarrese}, {Lucchin}, {Fontana}, \& {Giallongo}}]{Arnouts1999}
{Arnouts}, S., {Cristiani}, S., {Moscardini}, L., {et~al.} 1999, \mnras, 310,
  540

\bibitem[{{Baldry} {et~al.}(2006){Baldry}, {Balogh}, {Bower}, {Glazebrook},
  {Nichol}, {Bamford}, \& {Budavari}}]{Baldry2006}
{Baldry}, I.~K., {Balogh}, M.~L., {Bower}, R.~G., {et~al.} 2006, \mnras, 373,
  469

\bibitem[{{Baldry} {et~al.}(2004){Baldry}, {Glazebrook}, {Brinkmann},
  {Ivezi{\'c}}, {Lupton}, {Nichol}, \& {Szalay}}]{Baldry2004}
{Baldry}, I.~K., {Glazebrook}, K., {Brinkmann}, J., {et~al.} 2004, \apj, 600,
  681

\bibitem[{{Balogh} {et~al.}(2004){Balogh}, {Baldry}, {Nichol}, {Miller},
  {Bower}, \& {Glazebrook}}]{Balogh2004}
{Balogh}, M.~L., {Baldry}, I.~K., {Nichol}, R., {et~al.} 2004, \apjl, 615, L101

\bibitem[{{Balogh} {et~al.}(1999){Balogh}, {Morris}, {Yee}, {Carlberg}, \&
  {Ellingson}}]{Balogh1999}
{Balogh}, M.~L., {Morris}, S.~L., {Yee}, H.~K.~C., {Carlberg}, R.~G., \&
  {Ellingson}, E. 1999, \apj, 527, 54

\bibitem[{{Balogh} {et~al.}(2016){Balogh}, {McGee}, {Mok}, {Muzzin}, {van der
  Burg}, {Bower}, {Finoguenov}, {Hoekstra}, {Lidman}, {Mulchaey}, {Noble},
  {Parker}, {Tanaka}, {Wilman}, {Webb}, {Wilson}, \& {Yee}}]{Balogh2016}
{Balogh}, M.~L., {McGee}, S.~L., {Mok}, A., {et~al.} 2016, \mnras, 456, 4364

\bibitem[{{Bosch} {et~al.}(2017){Bosch}, {Armstrong}, {Bickerton}, {Furusawa},
  {Ikeda}, {Koike}, {Lupton}, {Mineo}, {Price}, {Takata}, {Tanaka}, {Yasuda},
  {AlSayyad}, {Becker}, {Coulton}, {Coupon}, {Garmilla}, {Huang}, {Krughoff},
  {Lang}, {Leauthaud}, {Lim}, {Lust}, {MacArthur}, {Mandelbaum}, {Miyatake},
  {Miyazaki}, {Murata}, {More}, {Okura}, {Owen}, {Swinbank}, {Strauss},
  {Yamada}, \& {Yamanoi}}]{Bosch2017temp}
{Bosch}, J., {Armstrong}, R., {Bickerton}, S., {et~al.} 2017, ArXiv e-prints,
  arXiv:1705.06766

\bibitem[{{Boselli} \& {Gavazzi}(2006)}]{Boselli2006}
{Boselli}, A., \& {Gavazzi}, G. 2006, \pasp, 118, 517

\bibitem[{{Brammer} {et~al.}(2008){Brammer}, {van Dokkum}, \&
  {Coppi}}]{Brammer2008}
{Brammer}, G.~B., {van Dokkum}, P.~G., \& {Coppi}, P. 2008, \apj, 686, 1503

\bibitem[{{Brodwin} {et~al.}(2013){Brodwin}, {Stanford}, {Gonzalez}, {Zeimann},
  {Snyder}, {Mancone}, {Pope}, {Eisenhardt}, {Stern}, {Alberts}, {Ashby},
  {Brown}, {Chary}, {Dey}, {Galametz}, {Gettings}, {Jannuzi}, {Miller},
  {Moustakas}, \& {Moustakas}}]{Brodwin2013}
{Brodwin}, M., {Stanford}, S.~A., {Gonzalez}, A.~H., {et~al.} 2013, \apj, 779,
  138

\bibitem[{{Bruzual} \& {Charlot}(2003)}]{Bruzual2003}
{Bruzual}, G., \& {Charlot}, S. 2003, \mnras, 344, 1000

\bibitem[{{Bufanda} {et~al.}(2017){Bufanda}, {Hollowood}, {Jeltema}, {Rykoff},
  {Rozo}, {Martini}, {Abbott}, {Abdalla}, {Allam}, {Banerji},
  {Benoit-L{\'e}vy}, {Bertin}, {Brooks}, {Carnero Rosell}, {Carrasco Kind},
  {Carretero}, {Cunha}, {da Costa}, {Desai}, {Diehl}, {Dietrich}, {Evrard},
  {Fausti Neto}, {Flaugher}, {Frieman}, {Gerdes}, {Goldstein}, {Gruen},
  {Gruendl}, {Gutierrez}, {Honscheid}, {James}, {Kuehn}, {Kuropatkin}, {Lima},
  {Maia}, {Marshall}, {Melchior}, {Miquel}, {Mohr}, {Ogando}, {Plazas},
  {Romer}, {Rooney}, {Sanchez}, {Santiago}, {Scarpine}, {Sevilla-Noarbe},
  {Smith}, {Soares-Santos}, {Sobreira}, {Suchyta}, {Tarle}, {Thomas}, {Tucker},
  {Walker}, \& {DES Collaboration}}]{Bufanda2017}
{Bufanda}, E., {Hollowood}, D., {Jeltema}, T.~E., {et~al.} 2017, \mnras, 465,
  2531

\bibitem[{{Butcher} \& {Oemler}(1984)}]{Butcher1984}
{Butcher}, H., \& {Oemler}, Jr., A. 1984, \apj, 285, 426

\bibitem[{{Calzetti} {et~al.}(2000){Calzetti}, {Armus}, {Bohlin}, {Kinney},
  {Koornneef}, \& {Storchi-Bergmann}}]{Calzetti2000}
{Calzetti}, D., {Armus}, L., {Bohlin}, R.~C., {et~al.} 2000, \apj, 533, 682

\bibitem[{{Chabrier}(2003)}]{Chabrier2003}
{Chabrier}, G. 2003, \pasp, 115, 763

\bibitem[{{Chung} {et~al.}(2011){Chung}, {Eisenhardt}, {Gonzalez}, {Stanford},
  {Brodwin}, {Stern}, \& {Jarrett}}]{Chung2011}
{Chung}, S.~M., {Eisenhardt}, P.~R., {Gonzalez}, A.~H., {et~al.} 2011, \apj,
  743, 34

\bibitem[{{Coil} {et~al.}(2011){Coil}, {Blanton}, {Burles}, {Cool},
  {Eisenstein}, {Moustakas}, {Wong}, {Zhu}, {Aird}, {Bernstein}, {Bolton}, \&
  {Hogg}}]{Coil2011}
{Coil}, A.~L., {Blanton}, M.~R., {Burles}, S.~M., {et~al.} 2011, \apj, 741, 8

\bibitem[{{Cowie} {et~al.}(1996){Cowie}, {Songaila}, {Hu}, \&
  {Cohen}}]{Cowie1996}
{Cowie}, L.~L., {Songaila}, A., {Hu}, E.~M., \& {Cohen}, J.~G. 1996, \aj, 112,
  839

\bibitem[{{D'Agostini}(2004)}]{DAgostini2004}
{D'Agostini}, G. 2004, ArXiv Physics e-prints, physics/0412069

\bibitem[{{Darvish} {et~al.}(2018){Darvish}, {Martin}, {Gon{\c c}alves},
  {Mobasher}, {Scoville}, \& {Sobral}}]{Darvish2018}
{Darvish}, B., {Martin}, C., {Gon{\c c}alves}, T.~S., {et~al.} 2018, \apj, 853,
  155

\bibitem[{{Darvish} {et~al.}(2016){Darvish}, {Mobasher}, {Sobral}, {Rettura},
  {Scoville}, {Faisst}, \& {Capak}}]{Darvish2016}
{Darvish}, B., {Mobasher}, B., {Sobral}, D., {et~al.} 2016, \apj, 825, 113

\bibitem[{{Davies} {et~al.}(2016){Davies}, {Robotham}, {Driver}, {Alpaslan},
  {Baldry}, {Bland-Hawthorn}, {Brough}, {Brown}, {Cluver}, {Holwerda},
  {Hopkins}, {Lara-L{\'o}pez}, {Mahajan}, {Moffett}, {Owers}, \&
  {Phillipps}}]{Davies2016}
{Davies}, L.~J.~M., {Robotham}, A.~S.~G., {Driver}, S.~P., {et~al.} 2016,
  \mnras, 455, 4013

\bibitem[{{Dekel} \& {Birnboim}(2006)}]{Dekel2006}
{Dekel}, A., \& {Birnboim}, Y. 2006, \mnras, 368, 2

\bibitem[{{Dekel} \& {Silk}(1986)}]{Dekel1986}
{Dekel}, A., \& {Silk}, J. 1986, \apj, 303, 39

\bibitem[{{Dressler}(1980)}]{Dressler1980}
{Dressler}, A. 1980, \apj, 236, 351

\bibitem[{{Fabian}(2012)}]{Fabian2012}
{Fabian}, A.~C. 2012, \araa, 50, 455

\bibitem[{{Fang} {et~al.}(2017){Fang}, {Faber}, {Koo}, {Rodriguez-Puebla},
  {Guo}, {Barro}, {Behroozi}, {Brammer}, {Chen}, {Dekel}, {Ferguson},
  {Gawiser}, {Giavalisco}, {Kartaltepe}, {Kocevski}, {Koekemoer}, {McGrath},
  {McIntosh}, {Newman}, {Pacifici}, {Pandya}, {Perez-Gonzalez}, {Primack},
  {Salmon}, {Trump}, {Weiner}, {Willner}, {Acquaviva}, {Dahlen}, {Finkelstein},
  {Finlator}, {Fontana}, {Galametz}, {Grogin}, {Gruetzbauch}, {Johnson},
  {Mobasher}, {Papovich}, {Pforr}, {Salvato}, {Santini}, {van der Wel},
  {Wiklind}, \& {Wuyts}}]{Fang2018temp}
{Fang}, J.~J., {Faber}, S.~M., {Koo}, D.~C., {et~al.} 2017, ArXiv e-prints,
  arXiv:1710.05489

\bibitem[{{Fazio} {et~al.}(2004){Fazio}, {Hora}, {Allen}, {Ashby}, {Barmby},
  {Deutsch}, {Huang}, {Kleiner}, {Marengo}, {Megeath}, {Melnick}, {Pahre},
  {Patten}, {Polizotti}, {Smith}, {Taylor}, {Wang}, {Willner}, {Hoffmann},
  {Pipher}, {Forrest}, {McMurty}, {McCreight}, {McKelvey}, {McMurray}, {Koch},
  {Moseley}, {Arendt}, {Mentzell}, {Marx}, {Losch}, {Mayman}, {Eichhorn},
  {Krebs}, {Jhabvala}, {Gezari}, {Fixsen}, {Flores}, {Shakoorzadeh}, {Jungo},
  {Hakun}, {Workman}, {Karpati}, {Kichak}, {Whitley}, {Mann}, {Tollestrup},
  {Eisenhardt}, {Stern}, {Gorjian}, {Bhattacharya}, {Carey}, {Nelson},
  {Glaccum}, {Lacy}, {Lowrance}, {Laine}, {Reach}, {Stauffer}, {Surace},
  {Wilson}, {Wright}, {Hoffman}, {Domingo}, \& {Cohen}}]{Fazio2004}
{Fazio}, G.~G., {Hora}, J.~L., {Allen}, L.~E., {et~al.} 2004, \apjs, 154, 10

\bibitem[{{Foltz} {et~al.}(2015){Foltz}, {Rettura}, {Wilson}, {van der Burg},
  {Muzzin}, {Lidman}, {Demarco}, {Nantais}, {DeGroot}, \& {Yee}}]{Foltz2015}
{Foltz}, R., {Rettura}, A., {Wilson}, G., {et~al.} 2015, \apj, 812, 138

\bibitem[{{Garilli} {et~al.}(2014){Garilli}, {Guzzo}, {Scodeggio},
  {Bolzonella}, {Abbas}, {Adami}, {Arnouts}, {Bel}, {Bottini}, {Branchini},
  {Cappi}, {Coupon}, {Cucciati}, {Davidzon}, {De Lucia}, {de la Torre},
  {Franzetti}, {Fritz}, {Fumana}, {Granett}, {Ilbert}, {Iovino}, {Krywult}, {Le
  Brun}, {Le F{\`e}vre}, {Maccagni}, {Ma{\l}ek}, {Marulli}, {McCracken},
  {Paioro}, {Polletta}, {Pollo}, {Schlagenhaufer}, {Tasca}, {Tojeiro},
  {Vergani}, {Zamorani}, {Zanichelli}, {Burden}, {Di Porto}, {Marchetti},
  {Marinoni}, {Mellier}, {Moscardini}, {Nichol}, {Peacock}, {Percival},
  {Phleps}, \& {Wolk}}]{Garilli2014}
{Garilli}, B., {Guzzo}, L., {Scodeggio}, M., {et~al.} 2014, \aap, 562, A23

\bibitem[{{Geha} {et~al.}(2012){Geha}, {Blanton}, {Yan}, \&
  {Tinker}}]{Geha2012}
{Geha}, M., {Blanton}, M.~R., {Yan}, R., \& {Tinker}, J.~L. 2012, \apj, 757, 85

\bibitem[{{Gladders} \& {Yee}(2000)}]{Gladders2000}
{Gladders}, M.~D., \& {Yee}, H.~K.~C. 2000, \aj, 120, 2148

\bibitem[{{Gladders} \& {Yee}(2005)}]{Gladders2005}
---. 2005, \apjs, 157, 1

\bibitem[{{Grazian} {et~al.}(2006){Grazian}, {Fontana}, {de Santis}, {Nonino},
  {Salimbeni}, {Giallongo}, {Cristiani}, {Gallozzi}, \&
  {Vanzella}}]{Grazian2006}
{Grazian}, A., {Fontana}, A., {de Santis}, C., {et~al.} 2006, \aap, 449, 951

\bibitem[{{Gunn} \& {Gott}(1972)}]{Gunn1972}
{Gunn}, J.~E., \& {Gott}, III, J.~R. 1972, \apj, 176, 1

\bibitem[{{Haines} {et~al.}(2015){Haines}, {Pereira}, {Smith}, {Egami},
  {Babul}, {Finoguenov}, {Ziparo}, {McGee}, {Rawle}, {Okabe}, \&
  {Moran}}]{Haines2015}
{Haines}, C.~P., {Pereira}, M.~J., {Smith}, G.~P., {et~al.} 2015, \apj, 806,
  101

\bibitem[{{Haines} {et~al.}(2017){Haines}, {Iovino}, {Krywult}, {Guzzo},
  {Davidzon}, {Bolzonella}, {Garilli}, {Scodeggio}, {Granett}, {de la Torre},
  {De Lucia}, {Abbas}, {Adami}, {Arnouts}, {Bottini}, {Cappi}, {Cucciati},
  {Franzetti}, {Fritz}, {Gargiulo}, {Le Brun}, {Le F{\`e}vre}, {Maccagni},
  {Ma{\l}ek}, {Marulli}, {Moutard}, {Polletta}, {Pollo}, {Tasca}, {Tojeiro},
  {Vergani}, {Zanichelli}, {Zamorani}, {Bel}, {Branchini}, {Coupon}, {Ilbert},
  {Moscardini}, {Peacock}, \& {Siudek}}]{Haines2017}
{Haines}, C.~P., {Iovino}, A., {Krywult}, J., {et~al.} 2017, \aap, 605, A4

\bibitem[{{Hennig} {et~al.}(2017){Hennig}, {Mohr}, {Zenteno}, {Desai},
  {Dietrich}, {Bocquet}, {Strazzullo}, {Saro}, {Abbott}, {Abdalla}, {Bayliss},
  {Benoit-L{\'e}vy}, {Bernstein}, {Bertin}, {Brooks}, {Capasso}, {Capozzi},
  {Carnero}, {Carrasco Kind}, {Carretero}, {Chiu}, {D'Andrea}, {daCosta},
  {Diehl}, {Doel}, {Eifler}, {Evrard}, {Fausti-Neto}, {Fosalba}, {Frieman},
  {Gangkofner}, {Gonzalez}, {Gruen}, {Gruendl}, {Gupta}, {Gutierrez},
  {Honscheid}, {Hlavacek-Larrondo}, {James}, {Kuehn}, {Kuropatkin}, {Lahav},
  {March}, {Marshall}, {Martini}, {McDonald}, {Melchior}, {Miller}, {Miquel},
  {Neilsen}, {Nord}, {Ogando}, {Plazas}, {Reichardt}, {Romer}, {Rozo},
  {Rykoff}, {Sanchez}, {Santiago}, {Schubnell}, {Sevilla-Noarbe}, {Smith},
  {Soares-Santos}, {Sobreira}, {Stalder}, {Stanford}, {Suchyta}, {Swanson},
  {Tarle}, {Thomas}, {Vikram}, {Walker}, \& {Zhang}}]{Hennig2017}
{Hennig}, C., {Mohr}, J.~J., {Zenteno}, A., {et~al.} 2017, \mnras, 467, 4015

\bibitem[{{Hilton} {et~al.}(2010){Hilton}, {Lloyd-Davies}, {Stanford}, {Stott},
  {Collins}, {Romer}, {Hosmer}, {Hoyle}, {Kay}, {Liddle}, {Mehrtens}, {Miller},
  {Sahl{\'e}n}, \& {Viana}}]{Hilton2010}
{Hilton}, M., {Lloyd-Davies}, E., {Stanford}, S.~A., {et~al.} 2010, \apj, 718,
  133

\bibitem[{{Hsieh} \& {Yee}(2014)}]{Hsieh2014}
{Hsieh}, B.~C., \& {Yee}, H.~K.~C. 2014, \apj, 792, 102

\bibitem[{{Ilbert} {et~al.}(2006){Ilbert}, {Arnouts}, {McCracken},
  {Bolzonella}, {Bertin}, {Le F{\`e}vre}, {Mellier}, {Zamorani}, {Pell{\`o}},
  {Iovino}, {Tresse}, {Le Brun}, {Bottini}, {Garilli}, {Maccagni}, {Picat},
  {Scaramella}, {Scodeggio}, {Vettolani}, {Zanichelli}, {Adami}, {Bardelli},
  {Cappi}, {Charlot}, {Ciliegi}, {Contini}, {Cucciati}, {Foucaud}, {Franzetti},
  {Gavignaud}, {Guzzo}, {Marano}, {Marinoni}, {Mazure}, {Meneux}, {Merighi},
  {Paltani}, {Pollo}, {Pozzetti}, {Radovich}, {Zucca}, {Bondi}, {Bongiorno},
  {Busarello}, {de La Torre}, {Gregorini}, {Lamareille}, {Mathez}, {Merluzzi},
  {Ripepi}, {Rizzo}, \& {Vergani}}]{Ilbert2006}
{Ilbert}, O., {Arnouts}, S., {McCracken}, H.~J., {et~al.} 2006, \aap, 457, 841

\bibitem[{{Ilbert} {et~al.}(2009){Ilbert}, {Capak}, {Salvato}, {Aussel},
  {McCracken}, {Sanders}, {Scoville}, {Kartaltepe}, {Arnouts}, {Le Floc'h},
  {Mobasher}, {Taniguchi}, {Lamareille}, {Leauthaud}, {Sasaki}, {Thompson},
  {Zamojski}, {Zamorani}, {Bardelli}, {Bolzonella}, {Bongiorno}, {Brusa},
  {Caputi}, {Carollo}, {Contini}, {Cook}, {Coppa}, {Cucciati}, {de la Torre},
  {de Ravel}, {Franzetti}, {Garilli}, {Hasinger}, {Iovino}, {Kampczyk},
  {Kneib}, {Knobel}, {Kovac}, {Le Borgne}, {Le Brun}, {Le F{\`e}vre}, {Lilly},
  {Looper}, {Maier}, {Mainieri}, {Mellier}, {Mignoli}, {Murayama}, {Pell{\`o}},
  {Peng}, {P{\'e}rez-Montero}, {Renzini}, {Ricciardelli}, {Schiminovich},
  {Scodeggio}, {Shioya}, {Silverman}, {Surace}, {Tanaka}, {Tasca}, {Tresse},
  {Vergani}, \& {Zucca}}]{Ilbert2009}
{Ilbert}, O., {Capak}, P., {Salvato}, M., {et~al.} 2009, \apj, 690, 1236

\bibitem[{{Jaff{\'e}} {et~al.}(2015){Jaff{\'e}}, {Smith}, {Candlish},
  {Poggianti}, {Sheen}, \& {Verheijen}}]{Jaffe2015}
{Jaff{\'e}}, Y.~L., {Smith}, R., {Candlish}, G.~N., {et~al.} 2015, \mnras, 448,
  1715

\bibitem[{{Jaff{\'e}} {et~al.}(2018){Jaff{\'e}}, {Poggianti}, {Moretti},
  {Gullieuszik}, {Smith}, {Vulcani}, {Fasano}, {Fritz}, {Tonnesen}, {Bettoni},
  {Hau}, {Biviano}, {Bellhouse}, \& {McGee}}]{Jaffe2018}
{Jaff{\'e}}, Y.~L., {Poggianti}, B.~M., {Moretti}, A., {et~al.} 2018, \mnras,
  476, 4753

\bibitem[{{Jian} {et~al.}(2017){Jian}, {Lin}, {Lin}, {Foucaud}, {Chen},
  {Chiueh}, {Bower}, {Cole}, {Chen}, {Burgett}, {Draper}, {Flewelling},
  {Huber}, {Kaiser}, {Kudritzki}, {Magnier}, {Metcalfe}, {Wainscoat}, \&
  {Waters}}]{Jian2017}
{Jian}, H.-Y., {Lin}, L., {Lin}, K.-Y., {et~al.} 2017, \apj, 845, 74

\bibitem[{{Jian} {et~al.}(2018){Jian}, {Lin}, {Oguri}, {Nishizawa}, {Takada},
  {More}, {Koyama}, {Tanaka}, \& {Komiyama}}]{Jian2018}
{Jian}, H.-Y., {Lin}, L., {Oguri}, M., {et~al.} 2018, \pasj, 70, S23

\bibitem[{{Kauffmann} {et~al.}(2004){Kauffmann}, {White}, {Heckman},
  {M{\'e}nard}, {Brinchmann}, {Charlot}, {Tremonti}, \&
  {Brinkmann}}]{Kauffmann2004}
{Kauffmann}, G., {White}, S.~D.~M., {Heckman}, T.~M., {et~al.} 2004, \mnras,
  353, 713

\bibitem[{{Kawinwanichakij} {et~al.}(2017){Kawinwanichakij}, {Papovich},
  {Quadri}, {Glazebrook}, {Kacprzak}, {Allen}, {Bell}, {Croton}, {Dekel},
  {Ferguson}, {Forrest}, {Grogin}, {Guo}, {Kocevski}, {Koekemoer}, {Labb{\'e}},
  {Lucas}, {Nanayakkara}, {Spitler}, {Straatman}, {Tran}, {Tomczak}, \& {van
  Dokkum}}]{Kawinwanichakij2017}
{Kawinwanichakij}, L., {Papovich}, C., {Quadri}, R.~F., {et~al.} 2017, \apj,
  847, 134

\bibitem[{{Kova{\v c}} {et~al.}(2014){Kova{\v c}}, {Lilly}, {Knobel},
  {Bschorr}, {Peng}, {Carollo}, {Contini}, {Kneib}, {Le F{\'e}vre}, {Mainieri},
  {Renzini}, {Scodeggio}, {Zamorani}, {Bardelli}, {Bolzonella}, {Bongiorno},
  {Caputi}, {Cucciati}, {de la Torre}, {de Ravel}, {Franzetti}, {Garilli},
  {Iovino}, {Kampczyk}, {Lamareille}, {Le Borgne}, {Le Brun}, {Maier},
  {Mignoli}, {Oesch}, {Pello}, {Montero}, {Presotto}, {Silverman}, {Tanaka},
  {Tasca}, {Tresse}, {Vergani}, {Zucca}, {Aussel}, {Koekemoer}, {Le Floc'h},
  {Moresco}, \& {Pozzetti}}]{Kovac2014}
{Kova{\v c}}, K., {Lilly}, S.~J., {Knobel}, C., {et~al.} 2014, \mnras, 438, 717

\bibitem[{{Lawrence} {et~al.}(2007){Lawrence}, {Warren}, {Almaini}, {Edge},
  {Hambly}, {Jameson}, {Lucas}, {Casali}, {Adamson}, {Dye}, {Emerson},
  {Foucaud}, {Hewett}, {Hirst}, {Hodgkin}, {Irwin}, {Lodieu}, {McMahon},
  {Simpson}, {Smail}, {Mortlock}, \& {Folger}}]{Lawrence2007}
{Lawrence}, A., {Warren}, S.~J., {Almaini}, O., {et~al.} 2007, \mnras, 379,
  1599

\bibitem[{{Le F{\`e}vre} {et~al.}(2013){Le F{\`e}vre}, {Cassata}, {Cucciati},
  {Garilli}, {Ilbert}, {Le Brun}, {Maccagni}, {Moreau}, {Scodeggio}, {Tresse},
  {Zamorani}, {Adami}, {Arnouts}, {Bardelli}, {Bolzonella}, {Bondi},
  {Bongiorno}, {Bottini}, {Cappi}, {Charlot}, {Ciliegi}, {Contini}, {de la
  Torre}, {Foucaud}, {Franzetti}, {Gavignaud}, {Guzzo}, {Iovino}, {Lemaux},
  {L{\'o}pez-Sanjuan}, {McCracken}, {Marano}, {Marinoni}, {Mazure}, {Mellier},
  {Merighi}, {Merluzzi}, {Paltani}, {Pell{\`o}}, {Pollo}, {Pozzetti},
  {Scaramella}, {Tasca}, {Vergani}, {Vettolani}, {Zanichelli}, \&
  {Zucca}}]{LeFevre2013}
{Le F{\`e}vre}, O., {Cassata}, P., {Cucciati}, O., {et~al.} 2013, \aap, 559,
  A14

\bibitem[{{Li} {et~al.}(2009){Li}, {Yee}, \& {Ellingson}}]{Li2009}
{Li}, I.~H., {Yee}, H.~K.~C., \& {Ellingson}, E. 2009, \apj, 698, 83

\bibitem[{{Li} {et~al.}(2012){Li}, {Yee}, {Hsieh}, \& {Gladders}}]{Li2012}
{Li}, I.~H., {Yee}, H.~K.~C., {Hsieh}, B.~C., \& {Gladders}, M. 2012, \apj,
  749, 150

\bibitem[{{Lin} {et~al.}(2014){Lin}, {Jian}, {Foucaud}, {Norberg}, {Bower},
  {Cole}, {Arnalte-Mur}, {Chen}, {Coupon}, {Hsieh}, {Heinis}, {Phleps}, {Chen},
  {Lee}, {Burgett}, {Chambers}, {Denneau}, {Draper}, {Flewelling}, {Hodapp},
  {Huber}, {Kaiser}, {Kudritzki}, {Magnier}, {Metcalfe}, {Price}, {Tonry},
  {Wainscoat}, \& {Waters}}]{Lin2014}
{Lin}, L., {Jian}, H.-Y., {Foucaud}, S., {et~al.} 2014, \apj, 782, 33

\bibitem[{{Lin} {et~al.}(2017){Lin}, {Hsieh}, {Lin}, {Oguri}, {Chen}, {Tanaka},
  {Chiu}, {Huang}, {Kodama}, {Leauthaud}, {More}, {Nishizawa}, {Bundy}, {Lin},
  \& {Miyazaki}}]{Lin2017temp}
{Lin}, Y.-T., {Hsieh}, B.-C., {Lin}, S.-C., {et~al.} 2017, ArXiv e-prints,
  arXiv:1709.04484

\bibitem[{{Loh} {et~al.}(2008){Loh}, {Ellingson}, {Yee}, {Gilbank}, {Gladders},
  \& {Barrientos}}]{Loh2008}
{Loh}, Y.-S., {Ellingson}, E., {Yee}, H.~K.~C., {et~al.} 2008, \apj, 680, 214

\bibitem[{{Lonsdale} {et~al.}(2003){Lonsdale}, {Smith}, {Rowan-Robinson},
  {Surace}, {Shupe}, {Xu}, {Oliver}, {Padgett}, {Fang}, {Conrow},
  {Franceschini}, {Gautier}, {Griffin}, {Hacking}, {Masci}, {Morrison},
  {O'Linger}, {Owen}, {P{\'e}rez-Fournon}, {Pierre}, {Puetter}, {Stacey},
  {Castro}, {Polletta}, {Farrah}, {Jarrett}, {Frayer}, {Siana}, {Babbedge},
  {Dye}, {Fox}, {Gonzalez-Solares}, {Salaman}, {Berta}, {Condon}, {Dole}, \&
  {Serjeant}}]{Lonsdale2003}
{Lonsdale}, C.~J., {Smith}, H.~E., {Rowan-Robinson}, M., {et~al.} 2003, \pasp,
  115, 897

\bibitem[{{Lupton} {et~al.}(2001){Lupton}, {Gunn}, {Ivezi{\'c}}, {Knapp}, \&
  {Kent}}]{Lupton2001}
{Lupton}, R., {Gunn}, J.~E., {Ivezi{\'c}}, Z., {Knapp}, G.~R., \& {Kent}, S.
  2001, in Astronomical Society of the Pacific Conference Series, Vol. 238,
  Astronomical Data Analysis Software and Systems X, ed. F.~R. {Harnden}, Jr.,
  F.~A. {Primini}, \& H.~E. {Payne}, 269

\bibitem[{{Maraston}(2005)}]{Maraston2005}
{Maraston}, C. 2005, \mnras, 362, 799

\bibitem[{{Marchesini} {et~al.}(2009){Marchesini}, {van Dokkum}, {F{\"o}rster
  Schreiber}, {Franx}, {Labb{\'e}}, \& {Wuyts}}]{Marchesini2009}
{Marchesini}, D., {van Dokkum}, P.~G., {F{\"o}rster Schreiber}, N.~M., {et~al.}
  2009, \apj, 701, 1765

\bibitem[{{Mauduit} {et~al.}(2012){Mauduit}, {Lacy}, {Farrah}, {Surace},
  {Jarvis}, {Oliver}, {Maraston}, {Vaccari}, {Marchetti}, {Zeimann},
  {Gonz{\'a}les-Solares}, {Pforr}, {Petric}, {Henriques}, {Thomas}, {Afonso},
  {Rettura}, {Wilson}, {Falder}, {Geach}, {Huynh}, {Norris}, {Seymour},
  {Richards}, {Stanford}, {Alexander}, {Becker}, {Best}, {Bizzocchi},
  {Bonfield}, {Castro}, {Cava}, {Chapman}, {Christopher}, {Clements}, {Covone},
  {Dubois}, {Dunlop}, {Dyke}, {Edge}, {Ferguson}, {Foucaud}, {Franceschini},
  {Gal}, {Grant}, {Grossi}, {Hatziminaoglou}, {Hickey}, {Hodge}, {Huang},
  {Ivison}, {Kim}, {LeFevre}, {Lehnert}, {Lonsdale}, {Lubin}, {McLure},
  {Messias}, {Mart{\'{\i}}nez-Sansigre}, {Mortier}, {Nielsen}, {Ouchi},
  {Parish}, {Perez-Fournon}, {Pierre}, {Rawlings}, {Readhead}, {Ridgway},
  {Rigopoulou}, {Romer}, {Rosebloom}, {Rottgering}, {Rowan-Robinson}, {Sajina},
  {Simpson}, {Smail}, {Squires}, {Stevens}, {Taylor}, {Trichas}, {Urrutia},
  {van Kampen}, {Verma}, \& {Xu}}]{Mauduit2012}
{Mauduit}, J.-C., {Lacy}, M., {Farrah}, D., {et~al.} 2012, \pasp, 124, 714

\bibitem[{{Maulbetsch} {et~al.}(2007){Maulbetsch}, {Avila-Reese},
  {Col{\'{\i}}n}, {Gottl{\"o}ber}, {Khalatyan}, \&
  {Steinmetz}}]{Maulbetsch2007}
{Maulbetsch}, C., {Avila-Reese}, V., {Col{\'{\i}}n}, P., {et~al.} 2007, \apj,
  654, 53

\bibitem[{{McDonald} {et~al.}(2016){McDonald}, {Stalder}, {Bayliss}, {Allen},
  {Applegate}, {Ashby}, {Bautz}, {Benson}, {Bleem}, {Brodwin}, {Carlstrom},
  {Chiu}, {Desai}, {Gonzalez}, {Hlavacek-Larrondo}, {Holzapfel}, {Marrone},
  {Miller}, {Reichardt}, {Saliwanchik}, {Saro}, {Schrabback}, {Stanford},
  {Stark}, {Vieira}, \& {Zenteno}}]{McDonald2016}
{McDonald}, M., {Stalder}, B., {Bayliss}, M., {et~al.} 2016, \apj, 817, 86

\bibitem[{{McGee} {et~al.}(2014){McGee}, {Bower}, \& {Balogh}}]{McGee2014}
{McGee}, S.~L., {Bower}, R.~G., \& {Balogh}, M.~L. 2014, \mnras, 442, L105

\bibitem[{{Merritt}(1984)}]{Merritt1984}
{Merritt}, D. 1984, \apj, 276, 26

\bibitem[{{Moore} {et~al.}(1996){Moore}, {Katz}, {Lake}, {Dressler}, \&
  {Oemler}}]{Moore1996}
{Moore}, B., {Katz}, N., {Lake}, G., {Dressler}, A., \& {Oemler}, A. 1996,
  \nat, 379, 613

\bibitem[{{Muldrew} {et~al.}(2012){Muldrew}, {Croton}, {Skibba}, {Pearce},
  {Ann}, {Baldry}, {Brough}, {Choi}, {Conselice}, {Cowan}, {Gallazzi}, {Gray},
  {Gr{\"u}tzbauch}, {Li}, {Park}, {Pilipenko}, {Podgorzec}, {Robotham},
  {Wilman}, {Yang}, {Zhang}, \& {Zibetti}}]{Muldrew2012}
{Muldrew}, S.~I., {Croton}, D.~J., {Skibba}, R.~A., {et~al.} 2012, \mnras, 419,
  2670

\bibitem[{{Muzzin} {et~al.}(2008){Muzzin}, {Wilson}, {Lacy}, {Yee}, \&
  {Stanford}}]{Muzzin2008}
{Muzzin}, A., {Wilson}, G., {Lacy}, M., {Yee}, H.~K.~C., \& {Stanford}, S.~A.
  2008, \apj, 686, 966

\bibitem[{{Muzzin} {et~al.}(2007){Muzzin}, {Yee}, {Hall}, \&
  {Lin}}]{Muzzin2007}
{Muzzin}, A., {Yee}, H.~K.~C., {Hall}, P.~B., \& {Lin}, H. 2007, \apj, 663, 150

\bibitem[{{Muzzin} {et~al.}(2009){Muzzin}, {Wilson}, {Yee}, {Hoekstra},
  {Gilbank}, {Surace}, {Lacy}, {Blindert}, {Majumdar}, {Demarco}, {Gardner},
  {Gladders}, \& {Lonsdale}}]{Muzzin2009}
{Muzzin}, A., {Wilson}, G., {Yee}, H.~K.~C., {et~al.} 2009, \apj, 698, 1934

\bibitem[{{Muzzin} {et~al.}(2012){Muzzin}, {Wilson}, {Yee}, {Gilbank},
  {Hoekstra}, {Demarco}, {Balogh}, {van Dokkum}, {Franx}, {Ellingson}, {Hicks},
  {Nantais}, {Noble}, {Lacy}, {Lidman}, {Rettura}, {Surace}, \&
  {Webb}}]{Muzzin2012}
---. 2012, \apj, 746, 188

\bibitem[{{Muzzin} {et~al.}(2013){Muzzin}, {Marchesini}, {Stefanon}, {Franx},
  {McCracken}, {Milvang-Jensen}, {Dunlop}, {Fynbo}, {Brammer}, {Labb{\'e}}, \&
  {van Dokkum}}]{Muzzin2013}
{Muzzin}, A., {Marchesini}, D., {Stefanon}, M., {et~al.} 2013, \apj, 777, 18

\bibitem[{{Muzzin} {et~al.}(2014){Muzzin}, {van der Burg}, {McGee}, {Balogh},
  {Franx}, {Hoekstra}, {Hudson}, {Noble}, {Taranu}, {Webb}, {Wilson}, \&
  {Yee}}]{Muzzin2014}
{Muzzin}, A., {van der Burg}, R.~F.~J., {McGee}, S.~L., {et~al.} 2014, \apj,
  796, 65

\bibitem[{{Nantais} {et~al.}(2017){Nantais}, {Muzzin}, {van der Burg},
  {Wilson}, {Lidman}, {Foltz}, {DeGroot}, {Noble}, {Cooper}, \&
  {Demarco}}]{Nantais2017}
{Nantais}, J.~B., {Muzzin}, A., {van der Burg}, R.~F.~J., {et~al.} 2017,
  \mnras, 465, L104

\bibitem[{{Noble} {et~al.}(2016){Noble}, {Webb}, {Yee}, {Muzzin}, {Wilson},
  {van der Burg}, {Balogh}, \& {Shupe}}]{Noble2016}
{Noble}, A.~G., {Webb}, T.~M.~A., {Yee}, H.~K.~C., {et~al.} 2016, \apj, 816, 48

\bibitem[{{Papovich} {et~al.}(2018){Papovich}, {Kawinwanichakij}, {Quadri},
  {Glazebrook}, {Labb{\'e}}, {Tran}, {Forrest}, {Kacprzak}, {Spitler},
  {Straatman}, \& {Tomczak}}]{Papovich2018}
{Papovich}, C., {Kawinwanichakij}, L., {Quadri}, R.~F., {et~al.} 2018, \apj,
  854, 30

\bibitem[{{Peng} {et~al.}(2010){Peng}, {Lilly}, {Kova{\v c}}, {Bolzonella},
  {Pozzetti}, {Renzini}, {Zamorani}, {Ilbert}, {Knobel}, {Iovino}, {Maier},
  {Cucciati}, {Tasca}, {Carollo}, {Silverman}, {Kampczyk}, {de Ravel},
  {Sanders}, {Scoville}, {Contini}, {Mainieri}, {Scodeggio}, {Kneib}, {Le
  F{\`e}vre}, {Bardelli}, {Bongiorno}, {Caputi}, {Coppa}, {de la Torre},
  {Franzetti}, {Garilli}, {Lamareille}, {Le Borgne}, {Le Brun}, {Mignoli},
  {Perez Montero}, {Pello}, {Ricciardelli}, {Tanaka}, {Tresse}, {Vergani},
  {Welikala}, {Zucca}, {Oesch}, {Abbas}, {Barnes}, {Bordoloi}, {Bottini},
  {Cappi}, {Cassata}, {Cimatti}, {Fumana}, {Hasinger}, {Koekemoer},
  {Leauthaud}, {Maccagni}, {Marinoni}, {McCracken}, {Memeo}, {Meneux}, {Nair},
  {Porciani}, {Presotto}, \& {Scaramella}}]{Peng2010}
{Peng}, Y.-j., {Lilly}, S.~J., {Kova{\v c}}, K., {et~al.} 2010, \apj, 721, 193

\bibitem[{{Phillips} {et~al.}(2014){Phillips}, {Wheeler}, {Boylan-Kolchin},
  {Bullock}, {Cooper}, \& {Tollerud}}]{Phillips2014}
{Phillips}, J.~I., {Wheeler}, C., {Boylan-Kolchin}, M., {et~al.} 2014, \mnras,
  437, 1930

\bibitem[{{Poggianti} {et~al.}(2006){Poggianti}, {von der Linden}, {De Lucia},
  {Desai}, {Simard}, {Halliday}, {Arag{\'o}n-Salamanca}, {Bower}, {Varela},
  {Best}, {Clowe}, {Dalcanton}, {Jablonka}, {Milvang-Jensen}, {Pello},
  {Rudnick}, {Saglia}, {White}, \& {Zaritsky}}]{Poggianti2006}
{Poggianti}, B.~M., {von der Linden}, A., {De Lucia}, G., {et~al.} 2006, \apj,
  642, 188

\bibitem[{{Quadri} {et~al.}(2012){Quadri}, {Williams}, {Franx}, \&
  {Hildebrandt}}]{Quadri2012}
{Quadri}, R.~F., {Williams}, R.~J., {Franx}, M., \& {Hildebrandt}, H. 2012,
  \apj, 744, 88

\bibitem[{{Raichoor} \& {Andreon}(2012)}]{Raichoor2012}
{Raichoor}, A., \& {Andreon}, S. 2012, \aap, 543, A19

\bibitem[{{Rawle} {et~al.}(2012){Rawle}, {Edge}, {Egami}, {Rex}, {Smith},
  {Altieri}, {Fiedler}, {Haines}, {Pereira}, {P{\'e}rez-Gonz{\'a}lez},
  {Portouw}, {Valtchanov}, {Walth}, {van der Werf}, \& {Zemcov}}]{Rawle2012}
{Rawle}, T.~D., {Edge}, A.~C., {Egami}, E., {et~al.} 2012, \apj, 747, 29

\bibitem[{{Rieke} {et~al.}(2004){Rieke}, {Young}, {Engelbracht}, {Kelly},
  {Low}, {Haller}, {Beeman}, {Gordon}, {Stansberry}, {Misselt}, {Cadien},
  {Morrison}, {Rivlis}, {Latter}, {Noriega-Crespo}, {Padgett}, {Stapelfeldt},
  {Hines}, {Egami}, {Muzerolle}, {Alonso-Herrero}, {Blaylock}, {Dole}, {Hinz},
  {Le Floc'h}, {Papovich}, {P{\'e}rez-Gonz{\'a}lez}, {Smith}, {Su}, {Bennett},
  {Frayer}, {Henderson}, {Lu}, {Masci}, {Pesenson}, {Rebull}, {Rho}, {Keene},
  {Stolovy}, {Wachter}, {Wheaton}, {Werner}, \& {Richards}}]{Rieke2004}
{Rieke}, G.~H., {Young}, E.~T., {Engelbracht}, C.~W., {et~al.} 2004, \apjs,
  154, 25

\bibitem[{{Saintonge} {et~al.}(2008){Saintonge}, {Tran}, \&
  {Holden}}]{Saintonge2008}
{Saintonge}, A., {Tran}, K.-V.~H., \& {Holden}, B.~P. 2008, \apjl, 685, L113

\bibitem[{{Santos} {et~al.}(2015){Santos}, {Altieri}, {Valtchanov}, {Nastasi},
  {B{\"o}hringer}, {Cresci}, {Elbaz}, {Fassbender}, {Rosati}, {Tozzi}, \&
  {Verdugo}}]{Santos2015}
{Santos}, J.~S., {Altieri}, B., {Valtchanov}, I., {et~al.} 2015, \mnras, 447,
  L65

\bibitem[{{Sobral} {et~al.}(2014){Sobral}, {Best}, {Smail}, {Mobasher},
  {Stott}, \& {Nisbet}}]{Sobral2014}
{Sobral}, D., {Best}, P.~N., {Smail}, I., {et~al.} 2014, \mnras, 437, 3516

\bibitem[{{Strateva} {et~al.}(2001){Strateva}, {Ivezi{\'c}}, {Knapp},
  {Narayanan}, {Strauss}, {Gunn}, {Lupton}, {Schlegel}, {Bahcall}, {Brinkmann},
  {Brunner}, {Budav{\'a}ri}, {Csabai}, {Castander}, {Doi}, {Fukugita}, {Gy{\H
  o}ry}, {Hamabe}, {Hennessy}, {Ichikawa}, {Kunszt}, {Lamb}, {McKay},
  {Okamura}, {Racusin}, {Sekiguchi}, {Schneider}, {Shimasaku}, \&
  {York}}]{Strateva2001}
{Strateva}, I., {Ivezi{\'c}}, {\v Z}., {Knapp}, G.~R., {et~al.} 2001, \aj, 122,
  1861

\bibitem[{{Tanaka} {et~al.}(2017){Tanaka}, {Coupon}, {Hsieh}, {Mineo},
  {Nishizawa}, {Speagle}, {Furusawa}, {Miyazaki}, \&
  {Murayama}}]{Tanaka2017temp}
{Tanaka}, M., {Coupon}, J., {Hsieh}, B.-C., {et~al.} 2017, ArXiv e-prints,
  arXiv:1704.05988

\bibitem[{{Tomczak} {et~al.}(2014){Tomczak}, {Quadri}, {Tran}, {Labb{\'e}},
  {Straatman}, {Papovich}, {Glazebrook}, {Allen}, {Brammer}, {Kacprzak},
  {Kawinwanichakij}, {Kelson}, {McCarthy}, {Mehrtens}, {Monson}, {Persson},
  {Spitler}, {Tilvi}, \& {van Dokkum}}]{Tomczak2014}
{Tomczak}, A.~R., {Quadri}, R.~F., {Tran}, K.-V.~H., {et~al.} 2014, \apj, 783,
  85

\bibitem[{{van den Bosch} {et~al.}(2008){van den Bosch}, {Aquino}, {Yang},
  {Mo}, {Pasquali}, {McIntosh}, {Weinmann}, \& {Kang}}]{vandenBosch2008}
{van den Bosch}, F.~C., {Aquino}, D., {Yang}, X., {et~al.} 2008, \mnras, 387,
  79

\bibitem[{{van der Burg} {et~al.}(2018){van der Burg}, {McGee}, {Aussel},
  {Dahle}, {Arnaud}, {Pratt}, \& {Muzzin}}]{vdBurg2018temp}
{van der Burg}, R.~F.~J., {McGee}, S., {Aussel}, H., {et~al.} 2018, ArXiv
  e-prints, arXiv:1807.00820

\bibitem[{{van der Burg} {et~al.}(2013){van der Burg}, {Muzzin}, {Hoekstra},
  {Lidman}, {Rettura}, {Wilson}, {Yee}, {Hildebrandt}, {Marchesini},
  {Stefanon}, {Demarco}, \& {Kuijken}}]{vanderBurg2013}
{van der Burg}, R.~F.~J., {Muzzin}, A., {Hoekstra}, H., {et~al.} 2013, \aap,
  557, A15

\bibitem[{{Veilleux} {et~al.}(2005){Veilleux}, {Cecil}, \&
  {Bland-Hawthorn}}]{Veilleux2005}
{Veilleux}, S., {Cecil}, G., \& {Bland-Hawthorn}, J. 2005, \araa, 43, 769

\bibitem[{{Vulcani} {et~al.}(2016){Vulcani}, {Treu}, {Schmidt}, {Morishita},
  {Dressler}, {Poggianti}, {Abramson}, {Brada{\v c}}, {Brammer}, {Hoag},
  {Malkan}, {Pentericci}, \& {Trenti}}]{Vulcani2016}
{Vulcani}, B., {Treu}, T., {Schmidt}, K.~B., {et~al.} 2016, \apj, 833, 178

\bibitem[{{Wagner} {et~al.}(2017){Wagner}, {Courteau}, {Brodwin}, {Stanford},
  {Snyder}, \& {Stern}}]{Wagner2017}
{Wagner}, C.~R., {Courteau}, S., {Brodwin}, M., {et~al.} 2017, \apj, 834, 53

\bibitem[{{Wang} {et~al.}(2016){Wang}, {Elbaz}, {Daddi}, {Finoguenov}, {Liu},
  {Schreiber}, {Mart{\'{\i}}n}, {Strazzullo}, {Valentino}, {van der Burg},
  {Zanella}, {Ciesla}, {Gobat}, {Le Brun}, {Pannella}, {Sargent}, {Shu}, {Tan},
  {Cappelluti}, \& {Li}}]{Wang2016}
{Wang}, T., {Elbaz}, D., {Daddi}, E., {et~al.} 2016, \apj, 828, 56

\bibitem[{{Webb} {et~al.}(2015){Webb}, {Noble}, {DeGroot}, {Wilson}, {Muzzin},
  {Bonaventura}, {Cooper}, {Delahaye}, {Foltz}, {Lidman}, {Surace}, {Yee},
  {Chapman}, {Dunne}, {Geach}, {Hayden}, {Hildebrandt}, {Huang}, {Pope},
  {Smith}, {Perlmutter}, \& {Tudorica}}]{Webb2015}
{Webb}, T., {Noble}, A., {DeGroot}, A., {et~al.} 2015, \apj, 809, 173

\bibitem[{{Weinzirl} {et~al.}(2017){Weinzirl}, {Arag{\'o}n-Salamanca}, {Gray},
  {Bamford}, {Rodr{\'{\i}}guez del Pino}, {Chies-Santos}, {B{\"o}hm}, {Wolf},
  \& {Cool}}]{Weinzirl2017}
{Weinzirl}, T., {Arag{\'o}n-Salamanca}, A., {Gray}, M.~E., {et~al.} 2017,
  \mnras, 471, 182

\bibitem[{{Wetzel} {et~al.}(2013){Wetzel}, {Tinker}, {Conroy}, \& {van den
  Bosch}}]{Wetzel2013}
{Wetzel}, A.~R., {Tinker}, J.~L., {Conroy}, C., \& {van den Bosch}, F.~C. 2013,
  \mnras, 432, 336

\bibitem[{{Wetzel} {et~al.}(2014){Wetzel}, {Tinker}, {Conroy}, \& {van den
  Bosch}}]{Wetzel2014}
---. 2014, \mnras, 439, 2687

\bibitem[{{Williams} {et~al.}(2009){Williams}, {Quadri}, {Franx}, {van Dokkum},
  \& {Labb{\'e}}}]{Williams2009}
{Williams}, R.~J., {Quadri}, R.~F., {Franx}, M., {van Dokkum}, P., \&
  {Labb{\'e}}, I. 2009, \apj, 691, 1879

\bibitem[{{Willis} {et~al.}(2018){Willis}, {Ramos-Ceja}, {Muzzin}, {Pacaud},
  {Yee}, \& {Wilson}}]{Willis2018temp}
{Willis}, J.~P., {Ramos-Ceja}, M.~E., {Muzzin}, A., {et~al.} 2018, \mnras,
  arXiv:1804.06475

\bibitem[{{Willis} {et~al.}(2013){Willis}, {Clerc}, {Bremer}, {Pierre},
  {Adami}, {Ilbert}, {Maughan}, {Maurogordato}, {Pacaud}, {Valtchanov},
  {Chiappetti}, {Thanjavur}, {Gwyn}, {Stanway}, \& {Winkworth}}]{Willis2013}
{Willis}, J.~P., {Clerc}, N., {Bremer}, M.~N., {et~al.} 2013, \mnras, 430, 134

\bibitem[{{Wilson} {et~al.}(2009){Wilson}, {Muzzin}, {Yee}, {Lacy}, {Surace},
  {Gilbank}, {Blindert}, {Hoekstra}, {Majumdar}, {Demarco}, {Gardner},
  {Gladders}, \& {Lonsdale}}]{Wilson2009}
{Wilson}, G., {Muzzin}, A., {Yee}, H.~K.~C., {et~al.} 2009, \apj, 698, 1943

\end{thebibliography}


\end{document}